\documentclass{ptephy_v1}

\preprintnumber{XXXX-XXXX} 

\usepackage{amssymb, amsmath, framed}
\usepackage{graphicx, color}

\usepackage{arydshln} 
\usepackage{ulem}

\newcommand{\beqa}{\begin{eqnarray}}
\newcommand{\eeqa}{\end{eqnarray}}

\newcommand{\cO}[1]{{\cal O}\left({#1}\right)}

\newcommand{\dbtilde}[1]{\tilde{\raisebox{0pt}[0.85\height]{$\tilde{#1}$}}}

\newcommand{\eq}[1]{(\ref{#1})}
\newcommand{\p}{\partial}

\newcommand{\mn}{\mathfrak{n}}
\newcommand{\ma}{\mathfrak{a}}

\allowdisplaybreaks[1]

\begin{document}

\title{Invertibility conditions for field transformations with derivatives: 
toward extensions of disformal transformation with higher derivatives}


\author{Eugeny Babichev}
\affil{Universit\'e Paris-Saclay, CNRS/IN2P3, IJCLab, 91405 Orsay, France}

\author[2,3]{Keisuke Izumi}
\affil{Kobayashi-Maskawa Institute, Nagoya University, Nagoya 464-8602, Japan}
\affil[3]{Department of Mathematics, Nagoya University, Nagoya 464-8602, Japan}

\author[4]{Norihiro Tanahashi}
\affil[4]{Department of Physics, Chuo University, 1-13-27 Kasuga, Bunkyo-ku, Tokyo 112-8551, Japan \email{tanahashi@phys.chuo-u.ac.jp}}

\author[5]{Masahide Yamaguchi}
\affil[5]{Department of Physics, Tokyo Institute of Technology, 2-12-1 Ookayama, Meguro-ku, Tokyo 152-8551, Japan}


\begin{abstract}%
We discuss a field transformation from 
fields $\psi_a$ to other
fields $\phi_i$ that involves derivatives, $\phi_i = \bar \phi_i(\psi_a, \partial_\alpha \psi_a, \ldots ;x^\mu)$,
and derive conditions for this transformation to be invertible, primarily focusing on the simplest case that the transformation maps between a pair of two fields and involves up to their first derivatives.
General field transformation of this type changes number of degrees of freedom, hence for the transformation to be  invertible, it must satisfy certain degeneracy conditions so that additional degrees of freedom do not appear.
Our derivation of necessary and sufficient conditions for invertible transformation is based on the method of characteristics, which is used to count the number of independent solutions of a given differential equation.
As applications of the invertibility conditions, we show some non-trivial examples of the invertible field transformations with derivatives, and also give a rigorous proof that a simple extension of the disformal transformation involving a second derivative of the scalar field is not invertible.
\end{abstract}

\subjectindex{A13, E03}

\maketitle

\section{Introduction}

Field transformations are quite ubiquitous in all of the fields of physics and mathematics. The reason is that by using suitable fields (variables), one can often get a better and more intuitive insight into physical phenomena, find a different form of equations of motion which may allow to obtain solutions more easily, and so on.
For these purposes, the (local) invertibility of a field transformation is 
essential because, otherwise, physics would not be the same after the field transformation. In case that a field transformation does not involve the derivatives, its local invertibility can be judged by the well-known inverse function theorem. 
On the other hand, when a field transformation does involve derivatives, it is clear that the invertibility conditions become much more complicated.
If one regards such a field transformation as differential equations for old variables (fields), one can naively expect the presence of integration constants associated with the derivatives, which 
breaks one-to-one correspondence between old and new variables. Thus, apparently, one might arrive at a conclusion that no invertible transformation with derivatives exists.  
But, of course, this is not true in general. 
If one assumes 
specific degeneracy of the derivative terms in a field transformation which 
prohibits the appearance of associated integration constants, one can have an invertible field transformation with the derivatives. 
Indeed, in our letter~\cite{Babichev:2019twf}  
we have explicitly given necessary and sufficient conditions for the invertibility of a field transformation involving two-fields and first derivatives.\footnote{See e.g.\ Ref.~\cite{Motohashi:2016prk} for earlier discussions on invertibility conditions of field transformations.}

The purpose of this paper is twofold. First, we give the full and complete proof of necessary and sufficient conditions for the invertibility of field transformations with derivatives. Here we fill some gaps of our proof that has been omitted in our letter~\cite{Babichev:2019twf}. More importantly, we provide the full proof for sufficient condition, while in the letter we gave it only within the perturbative regime.
The second goal 
of this paper is to prove the no-go theorem of disformal transformation of the metric that involves second derivatives of a scalar field.
We are strongly motivated by pursuing the extensions of conformal and disformal transformations, which are often used in gravity, cosmology, and many other fields (see e.g.\ Ref.~\cite{Achour:2016rkg} for classification of new theories generated from a simpler theory~\cite{Horndeski:1974wa} using disformal transformations).
A disformal transformation involving the first derivative of a scalar field is a natural extension of a conformal transformation~\cite{Bekenstein:1992pj}.
The next natural question is whether one can further extend a disformal transformation to that involving the second derivatives of a field or even its higher derivatives.
As a useful application of our invertibility conditions, we explicitly prove that there is no invertible disformal transformation involving the second derivatives
given by $\tilde g = C(\chi,X)g + D(\chi,X)\nabla\chi\nabla\chi + E(\chi,X)\nabla\nabla\chi~(X\equiv(\nabla\chi)^2)$ with $E\neq 0$. To the best of our knowledge, this is the first rigorous proof of the absence of such a transformation.

The organization of this paper is as follows. In section~\ref{sec:necessary}, we give a complete derivation of the necessary conditions.
The results in sections \ref{sec:necessary_general} apply to field transformations involving arbitrary number of fields and their first derivatives. To show the explicit form of the invertibility conditions, in section~\ref{sec:two-fields} and \ref{sec:2sim} we focus on 
the field transformation between two fields and their first derivatives. 
In the section~\ref{Sec:Suf}, the complete proof of the sufficient conditions is given.
As an application of our result,
in section~\ref{sec:examples} we construct some examples of invertible field transformation by solving the invertibility conditions.
As another application, in section~\ref{sec:no-go}, the no-go theorem for a class of disformal transformations of the metric with second derivatives is proven. Section~\ref{sec:discussions} is devoted to conclusions and discussion.
In Appendix~\ref{App:field-number}, we examine a field transformation that changes the number of fields and show that it cannot be invertible in our sense.
In Appendices~\ref{App:derivingformula} and \ref{App:longcalculation}, we show details of some calculations in section~\ref{sec:necessary}.
Appendix~\ref{App:one-field} shows that transformations between single fields and their derivatives can never be invertible, which implies that the two-field case we focus on is the simplest nontrivial case. 
In Appendix~\ref{App:implicit-function}, the necessity for nonlinear analysis in the inverse function theorem is explained.
Appendix~\ref{App:Setting-U} gives some technical details of the derivation in section~\ref{sec:examples},
and Appendix~\ref{app:gendisformal}  discusses an extension of
the results given in section~\ref{sec:no-go}.
\\

{\it Notations.}
Here we summarize the convention for indices used in this work.
\begin{align}
\partial_{\mu_1\mu_2\dots \mu_n} &:= \partial_{\mu_1}\partial_{\mu_2}\dots\partial_{\mu_n}~, \\
C^{(\mu_1\mu_2)}&=\frac12 \left(  C^{\mu_1\mu_2} + C^{\mu_2\mu_1} \right), \\
C^{[\mu_1\mu_2]}&=\frac12 \left(  C^{\mu_1\mu_2} - C^{\mu_2\mu_1} \right).
\end{align}
$C^{(\mu_1\mu_2\dots \mu_n)}$ and $C^{[\mu_1\mu_2\dots \mu_n]}$ are defined similarly by permutation (with a factor $(n!)^{-1}$).
We introduce the Levi-Civita symbol $\epsilon^{i_1\dots i_n i_{n+1} \dots i_{N}}$, which satisfies the following identities;
\begin{eqnarray}
&&\epsilon^{i_1\dots i_n i_{n+1} \dots i_{N}}\epsilon_{j_1\dots j_n j_{n+1}\dots j_{N}} M_{i_1}{}^{j_1} \dots M_{i_n}{}^{j_n} M_{i_{n+1}}{}^{k}  \nonumber \\
&&\qquad
=\epsilon^{i_1\dots  i_{N}}\epsilon_{j_1\dots j_{N}}  
\delta^{j_1}_{[l_1}\dots\delta^{j_n}_{l_n} \delta^{k}_{l_{n+1}]}  M_{i_1}{}^{l_1}\dots M_{i_{n+1}}{}^{l_{n+1}}\nonumber \\
&&\qquad
=\frac{1}{(n+1)!(N-n-1)!}\epsilon^{i_1\dots  i_{N}}\epsilon_{j_1\dots j_{N}}\epsilon^{j_1\dots  j_n k l_{n+2}\dots l_N}\epsilon_{l_1\dots l_{N}}    M_{i_1}{}^{l_1}\dots M_{i_{n+1}}{}^{l_{n+1}}\nonumber \\
&&\qquad
=\frac{n!(N-n)!}{(n+1)!(N-n-1)!}\epsilon^{i_1\dots  i_{N}}
\epsilon_{l_1\dots l_{N}}    
M_{i_1}{}^{l_1}\dots M_{i_{n+1}}{}^{l_{n+1}}\delta^{k}_{[j_{n+1}}\delta^{l_{n+2}}_{j_{n+2}}\dots\delta^{l_N}_{j_N]}
\nonumber \\ &&\qquad
=
\frac{N-n}{n+1}
\epsilon^{i_1\dots  i_{N}}
\epsilon_{l_1\dots l_{N}}    
M_{i_1}{}^{l_1}\dots M_{i_{n+1}}{}^{l_{n+1}}
\delta^{k}_{[j_{n+1}}\delta^{l_{n+2}}_{j_{n+2}}\dots\delta^{l_N}_{j_N]}~. 
\label{eeM...M}
\end{eqnarray}
Throughout this work, we do not distinguish the lower and upper indices for the field space indices $a,b,\ldots$ and $i,j,\dots$,
while in some parts upper/lower indices are used for clarity of the notation.

\section{Derivation of the necessary conditions}
\label{sec:necessary}

In this section we present a complete derivation of necessary conditions for invertibility of the transformation  $\phi_i = \bar \phi_i(\psi_a, \partial_\alpha \psi_a, x^\mu)$ that transforms fields $\phi_i$ into fields $\psi_a$.\footnote{Here we concentrate on the case with first-order derivatives. However, our idea of the use of characteristics and degeneracy applies to the case with arbitrary order derivatives (and arbitrary number of fields) in the same way.}
We divide the derivation of the invertibility conditions into two parts.
In the section~\ref{sec:necessary_general}, we derive necessary conditions for the invertibility. An invertible transformation preserves the number of degree of freedom, while a transformation with field derivatives typically generates additional degrees of freedom. To formulate such an idea mathematically, we employ the method of characteristics for a differential equation to count the number of propagating modes~\cite{CH:1962}.

In section~\ref{sec:degeneracies} we explain our approach to derive the necessary conditions for the invertibility based on the method of characteristics. 
In this approach, the transformation equation~(\ref{redef}) given below is converted to a set of differential equations that relates old variables to new ones, and the number of independent solutions for these equations is related to the number of degrees of freedom.
To establish the invertibility, 
the transformation must satisfy certain degeneracy conditions to remove unnecessary additional modes originating from the derivatives in the transformation equation.
Such degeneracy conditions must be imposed at each order of the aforementioned differential equations, which are summarized in sections~\ref{sec:leading} and \ref{sec:subleading}.
This procedure should be applied iteratively until the number of independent solutions are reduced appropriately, and then we may impose the non-degeneracy conditions to ensure the number of degree of freedom is not changed by the transformation.
We summarize this procedure in section~\ref{sec:subsubleading}.

The procedure in \ref{sec:necessary_general} applies to field transformation for general number of fields. For illustration of our method, we focus on the transformation between two fields in section~\ref{sec:two-fields} and \ref{sec:2sim}.
The necessary conditions for the invertibility in the two-field case are derived in section~\ref{sec:two-fields} based on the general method introduced in the previous sections.
The expressions of these conditions are rather complicated, and actually they can be simplified by solving a part of the conditions explicitly. Based on such an idea, the expressions of the necessary conditions are simplified in sections~\ref{sec:2sim_1}, \ref{sec:2sim_2} and \ref{sec:2simfin}. After simplification, the necessary conditions for the invertibility of a two-field transformations are summarized in Eqs.~(\ref{aVU}), (\ref{nBm}) and (\ref{nBU}). 

After deriving the necessary conditions in this section, in section \ref{Sec:Suf} we show that these conditions are actually sufficient to guarantee the invertibility. 

\subsection{
Method of characteristics as the key to derive the necessary conditions}
\label{sec:necessary_general}

\subsubsection{
Invertibility and number of degeneracies}
\label{sec:degeneracies}

Let us consider a field transformation from $\psi_a$ to $\phi_i$ given by
\begin{equation}
\phi_i = \bar \phi_i(\psi_a, \partial_\alpha \psi_a, x^\mu) ,\label{redef}
\end{equation}
where $\bar \phi_i$ is a function of $\psi_a$, $\partial_\alpha \psi_a$ and $x^\mu$.
We suppose that
the numbers of fields before and after transformation are the same, 
that is, we assume that both $a$ and $i$ run from $1$ to $N$, where $N$ is the number of field $\psi_a$.\footnote{For a field transformation that involves two fields and its first derivatives, it can be explicitly shown that a transformation that changes the number of fields can never be invertible. See Appendix~\ref{App:field-number} for details.}

If the transformation (\ref{redef}) is invertible, it does not change physical properties of theories before and after the transformation, and particularly the causal structure should be invariant.
If the transformation changes the number of characteristic hypersurfaces, 
the causal structure is changed correspondingly. 
Therefore, for the invertibility, the transformation should not change the number of characteristic hypersurfaces.
This gives the necessary conditions for invertibility.

To examine if the transformation (\ref{redef}) changes the number of characteristic hypersurfaces, we employ the method of characteristics for partial differential equations.
This method can be applied only to quasi-linear differential equations, while the transformation equation~(\ref{redef}) is nonlinear in $\partial_\alpha\psi_i$ in general.
To convert Eq.~\eq{redef} to a quasi-linear partial differential equation, 
we act a differential operator $K^{(\mu_1\dots\mu_n)}_{bi}\partial_{\mu_1\dots \mu_n}$ on it to obtain
\begin{align}
&K^{(\mu_1\dots\mu_n)}_{bi}\partial_{\mu_1\dots \mu_n} \phi_i
 \nonumber \\
&
=K^{(\mu_1\dots\mu_n)}_{bi}A^{\mu_{n+1}}_{ia}\partial_{\mu_1\dots \mu_{n+1}} \psi_a  \nonumber \\
&\quad
+\left(K^{(\mu_1\dots\mu_n)}_{bi}B_{ia}+ n\, K^{(\alpha\mu_1\dots\mu_{n-1})}_{bi} \partial_\alpha A^{\mu_n}_{ia}
\right)\partial_{\mu_1\dots \mu_n} \psi_a 
 \nonumber \\
&\quad
+\left(n\,K^{(\alpha \mu_1\dots\mu_{n-1})}_{bi} \partial_\alpha B_{ia}+ \frac{n(n-1)}{2}K^{(\alpha_1\alpha_2\mu_1\dots\mu_{n-2})}_{bi} \partial_{\alpha_1\alpha_2} A^{\mu_{n-1}}_{ia}\right)\partial_{\mu_1\dots \mu_{n-1}} \psi_a + \cdots 
 \nonumber \\
&\quad
+\left( \binom{n}{k} K^{(\alpha_1 \dots \alpha_k \mu_1\dots\mu_{n-k})}_{bi} \partial_{\alpha_1 \dots \alpha_k} B_{ia}
+ \binom{n}{k+1} K^{(\alpha_1 \dots \alpha_{k+1}\mu_1\dots\mu_{n-k-1})}_{bi} \partial_{\alpha_1 \dots \alpha_{k+1}} A^{\mu_{n-k}}_{ia}\right)\partial_{\mu_1\dots \mu_{n-k}} \psi_a 
 \nonumber \\
&\quad
+ \cdots 
+ \cO{\partial^{\lfloor\frac{n}{2}\rfloor+1}\psi_a,\partial^{\lfloor\frac{n}{2}\rfloor}\psi_a,\ldots} 
~,
\label{eq:C}
\end{align}
where $A^\alpha_{ia}, B_{ia} $ are $N\times N$ matrices defined by
\begin{equation}
A^\alpha_{ia} := \frac{\partial  \bar  \phi_i}{\partial(\partial_\alpha \psi_a)}~,
\qquad
B_{ia} := \frac{\partial  \bar  \phi_i}{\partial \psi_a} ~,
\end{equation}
and $\binom{n}{k}$ is the binomial coefficient
for $k\leq n - \lfloor \frac{n}{2}\rfloor -2 $.
In this expression, $\partial^{\lfloor\frac{n}{2}\rfloor+2}\psi_a$ and higher derivatives of $\psi_a$ appear linearly, while the $\cO{\partial^{\lfloor\frac{n}{2}\rfloor+1}\psi_a,\partial^{\lfloor\frac{n}{2}\rfloor}\psi_a,\ldots}$ part is a nonlinear function of lower-order derivatives of $\psi_a$.
We show the derivation of the formula (\ref{eq:C}) in Appendix~\ref{App:derivingformula}.

The operation of the differential operator $K^{(\mu_1\dots\mu_n)}_{bi}\partial_{\mu_1\dots \mu_n}$ 
introduces additional characteristic hypersurfaces
to those of Eq.~\eq{redef}. 
Hence, the characteristics of $\psi_a$, determined by the right-hand side of Eq.~\eq{eq:C}, are 
comprised of the original ones inherent to the field transformation equation \eq{redef} and the additional ones generated by
the operation of $K^{(\mu_1\dots\mu_n)}_{bi}\partial_{\mu_1\dots \mu_n}$.

Now we can analyse the characteristics of quasi-linear differential equation \eq{eq:C},
regarding this equation as partial differential equations on $\psi_a$.
We have $N$ differential equations with $(n+1)$-th order derivatives of $\psi_a$.
Therefore, they generically give $N\times(n+1)$ integration constants in a solution $\psi_a$, which corresponds 
$N\times(n+1)$ characteristics. 
However, if the field transformation~(\ref{redef}) is invertible, there are no characteristics 
inherent to (\ref{redef}), which implies that Eq.~(\ref{eq:C}) has only the additional characteristics generated by the operation of $K^{(\mu_1\dots\mu_n)}_{bi}\partial_{\mu_1\dots \mu_n}$.
The number of these additional characteristics is $N\times n$, 
hence there is a mismatch between the number of derivatives in the equations, $N\times(n+1)$, and that of the characteristics required by the invertibility, $N\times n$. 

Such a mismatch can be resolved if the structure of the highest-order derivative part is degenerate. 
Since the difference between them is $N$, the invertibility requires $N$ degrees of degeneracies.
We derive the conditions giving such $N$ degeneracies below.

\subsubsection{Degeneracy condition at leading order}
\label{sec:leading}

The characteristic equation for Eq.~\eq{eq:C} is given by
\begin{equation}
\det\left(K^{(\mu_1\dots\mu_n)}_{bi}A^{\mu_{n+1}}_{ia}\xi_{\mu_1}\dots \xi_{\mu_{n+1}}\right)=
\det\left(K^{(\mu_1\dots\mu_n)}_{bi}\xi_{\mu_1}\dots \xi_{\mu_{n}}\right) \det \left(A^{\mu_{n+1}}_{ia} \xi_{\mu_{n+1}}\right) =0,
\label{leading_char}
\end{equation}
where $\xi_\mu$ is a vector which is not tangent to would-be characteristics hypersurfaces. 
If $\det \left(A^{\mu}_{ia} \xi_{\mu}\right)$ does not vanish identically for any $\xi_\mu$, the characteristic equation (\ref{leading_char}) implies that there are $N \times (n+1)$ characteristics, and then the transformation (\ref{redef}) is not invertible as explained above.
Hence, for the invertibility $A^{\mu}_{ia} \xi_\mu$ must be degenerate for any $\xi_\mu$, that is, 
\begin{equation}
\det \left(A^\mu_{ia} \xi_\mu\right) =0 \qquad {\mbox{for any $\xi_\mu$}} .
\end{equation}
This condition is equivalent to 
\begin{equation}
\sum_{a \in S_n} \mbox{sgn} (a)  A^{(\alpha_1}_{1 a_1}\dots A^{\alpha_{N})}_{N a_{N}}
=\frac{1}{N!}\epsilon^{i_1\dots i_N} \epsilon^{a_1\dots a_N} A^{\alpha_1}_{i_1a_1}\dots A^{\alpha_N}_{i_N a_N}
=0~,
\label{LO}
\end{equation}
where $N$ is the number of fields 
and the sum is computed over the set $S_n$ of all permutations $a=\{a_1,\dots,a_N\}$ of $\{1,\dots, N\}$. 
Here, $\mathrm{sgn}(a)$ denotes the signature of a permutation $a$, which is $+1$ whenever the reordering $a$ can be achieved by successively interchanging two entries an even number of times, and $-1$ whenever it can be achieved by an odd number of such interchanges.

\subsubsection{Degeneracy condition at subleading order}
\label{sec:subleading}

The condition \eq{LO} indicates the degeneracy of the highest-order derivatives of Eq.~\eq{eq:C}. 
If the number of degeneracy is $N$ ({\it i.e.} the dimension of kernel $A^\mu_{ia} \xi_\mu$ is $N$), it implies that $A^\mu_{ia}=0$, and thus transformation is independent of $\p_\mu \psi_a$. 
It is a trivial case; we can directly use the implicit function theorem to Eq.~\eq{redef}. 
Therefore, we consider non-trivial cases where  the number of degeneracy of \eq{LO} is less than $N$. 
Here, we discuss the case where the number of degeneracy of \eq{LO} is $1$ for simplicity.
Generic cases may be discussed in a similar manner.

For later use, we introduce the adjugate matrix of $A^\mu_{ia}$, which is the transposed cofactor matrix, as
\begin{align}
{\bar A}^{\alpha_1\dots\alpha_{N-1}}_{ai}:={}^t{\bar A}^{\alpha_1\dots\alpha_{N-1}}_{ia}  &=  (-1)^{i+a} \sum_{\bar a \in \bar S_n}  \mbox{sgn} (\bar a) A^{(\alpha_1}_{\bar i_1 \bar a_1}\dots A^{\alpha_{N-1})}_{\bar i_{N-1} \bar a_{N-1}} \nonumber\\
&= \frac{1}{(N-1)!}
\epsilon_i^{~i_1\dots i_{N-1}}
\epsilon_a^{~ a_1\dots a_{N-1}} A^{\alpha_1}_{i_1a_1}\dots A^{\alpha_{N-1}}_{i_{N-1} a_{N-1}}
~,
\label{defCF}
\end{align}
where $\{\bar i_1,\dots,\bar i_{N-1}\}=\{1,\dots, i-1,i+1, \dots,N\}$ and the sum is computed over the set $\bar S_n$ of all permutations $\bar a = \{\bar a_1,\dots \bar a_{N-1}\}$ of $\{ 1,\dots,a-1,a+1,\dots,N \}$. 
This adjugate matrix satisfies
\begin{equation}
{\bar A}^{(\alpha_1\dots\alpha_{N-1}}_{ai} A^{\alpha_N)}_{ib}= 0~,
\qquad
A^{(\alpha_1}_{ia}{\bar A}^{\alpha_2\dots\alpha_N)}_{aj} =0~. 
\label{AA} 
\end{equation}
As commented above, we suppose that the dimension of kernel of $A^\mu_{ia} \xi_\mu$ is $1$ for any $\xi_\mu$. 
Then the adjugate matrix 
is a rank-1 matrix.

The fact that the highest derivative part of Eq.~\eq{eq:C} 
is degenerate in one dimension implies that Eq.~\eq{eq:C} contains one equation that involves only lower-order derivatives of $\psi_a$. This does not necessarily imply that one of the equations in \eq{eq:C} contains only lower derivative terms, instead, generically, by combining equations in~\eq{eq:C} one should be able to find one lower-derivative equation.
Let us extract this subleading equation. 
First, setting the operator $K_{bi}^{(\mu_1\ldots \mu_n)}\partial_{\mu_1\ldots\mu_n}$ in Eq.~\eq{eq:C} to ${\tilde K}^{(\mu_1\dots\mu_m)}_{b} {\bar A}^{\alpha_1\dots\alpha_{N-1}}_{bi} \partial_{\mu_1\dots\mu_m\alpha_1\dots\alpha_{N-1}} $ with $m=n-N+1$,
Eq.~\eq{eq:C} yields
\begin{eqnarray}
&&{\tilde K}^{(\mu_1\dots\mu_m)}_{b} {\bar A}^{\alpha_1\dots\alpha_{N-1}}_{bi} \partial_{\mu_1\dots\mu_m\alpha_1\dots\alpha_{N-1}} \phi_i  \nonumber \\
&&=
{\tilde K}^{(\mu_1\dots\mu_m)}_{b} {\bar A}^{(\alpha_1\dots\alpha_{N-1}}_{bi}A^{\alpha_N)}_{ia} \partial_{\mu_1\dots\mu_m\alpha_1\dots\alpha_{N}} \psi_a \nonumber \\
&& \quad
+{\tilde K}^{(\mu_1\dots\mu_m)}_{b} \left[  {\bar A}^{\alpha_1\dots\alpha_{N-1}}_{bi} B_{ia} + (N-1)  {\bar A}^{\beta\alpha_1\dots\alpha_{N-2}}_{bi}\left(\partial_\beta A^{\alpha_{N-1}}_{ia}\right) \right] \partial_{\mu_1\dots\mu_m\alpha_1\dots\alpha_{N-1}} \psi_a \nonumber \\
&& \quad
+m{\tilde K}^{(\beta\mu_1\dots\mu_{m-1})}_{b} {\bar A}^{\alpha_1\dots\alpha_{N-1}}_{bi} \left( \partial_\beta A^{\mu_m}_{ia} \right)\partial_{\mu_1\dots\mu_m\alpha_1\dots\alpha_{N-1}} \psi_a 
+ \cO { \partial^{m+N-2} \psi } .
\label{E1}
\end{eqnarray}
The first term of the right-hand side vanishes because of Eq.~(\ref{AA}).
Moreover, equation (\ref{AA}) gives 
\begin{equation}
{\bar A}^{(\alpha_1\dots\alpha_{N-1}}_{bi} \left(\partial_\beta A^{\alpha_N)}_{ia} \right) =- \left( \partial_\beta {\bar A}^{(\alpha_1\dots\alpha_{N-1}}_{bi} \right)A^{\alpha_N)}_{ia}.
\label{AbardA=0}
\end{equation}
This equation shows that the last term on the right-hand side of Eq.~(\ref{E1}) is written as 
\begin{eqnarray}
&&m{\tilde K}^{(\beta\mu_1\dots\mu_{m-1})}_{b} {\bar A}^{\alpha_1\dots\alpha_{N-1}}_{bi} \left( \partial_\beta A^{\mu_m}_{ia} \right)\partial_{\mu_1\dots\mu_m\alpha_1\dots\alpha_{N-1}} \psi_a  \nonumber \\
&&\quad
=-m{\tilde K}^{(\beta\mu_1\dots\mu_{m-1})}_{b}\left( \partial_\beta {\bar A}^{\alpha_1\dots\alpha_{N-1}}_{bi} \right)  A^{\mu_m}_{ia} \partial_{\mu_1\dots\mu_m\alpha_1\dots\alpha_{N-1}} \psi_a .
\end{eqnarray}
Next,
Eq.~(\ref{eq:C}) with
$K^{(\mu_1\dots\mu_n)}_{bi}\partial_{\mu_1\dots \mu_n}$ replaced by
$m{\tilde K}^{(\beta\mu_1\dots\mu_{m-1})}_{b} \left( \partial_\beta {\bar A}^{\alpha_1\dots\alpha_{N-1}}_{bi}  \right) \partial_{\mu_1\dots\mu_{m-1}\alpha_1\dots\alpha_{N-1}} $
gives 
\begin{eqnarray}
&&m{\tilde K}^{(\beta\mu_1\dots\mu_{m-1})}_{b} \left( \partial_\beta {\bar A}^{\alpha_1\dots\alpha_{N-1}}_{bi} \right)  \partial_{\mu_1\dots\mu_{m-1}\alpha_1\dots\alpha_{N-1}} \phi_i \nonumber \\
&&\quad
=m{\tilde K}^{(\beta\mu_1\dots\mu_{m-1})}_{b} \left( \partial_\beta {\bar A}^{\alpha_1\dots\alpha_{N-1}}_{bi} \right) A^{\mu_m}_{ia} \partial_{\mu_1\dots\mu_m\alpha_1\dots\alpha_{N-1}} \psi_a  
+\cO{\partial^{m+N-2}\psi}.
\end{eqnarray}
Then, the last term on the right-hand side of Eq.~(\ref{E1}) can be replaced with the derivatives of $\phi_i$ and $\cO{\partial^{m+N-2}\psi}$ terms. 
Expanding Eq.~(\ref{E1}) up to $\partial^{m+N-2} \psi$ and eliminating some 
$\partial^{m+N-1}\psi$ and $\partial^{m+N-2}\psi$
terms 
in favor of $\partial^{m+N-2}\phi$ and $\partial^{m+N-3}\phi$ terms as we did above,
we finally obtain
\begin{eqnarray}
&&{\tilde K}^{(\mu_1\dots\mu_m)}_{b} {\bar A}^{\alpha_1\dots\alpha_{N-1}}_{bi} \partial_{\mu_1\dots\mu_m\alpha_1\dots\alpha_{N-1}} \phi_i   
+ m{\tilde K}^{(\beta\mu_1\dots\mu_{m-1})}_{b} \left( \partial_\beta {\bar A}^{\alpha_1\dots\alpha_{N-1}}_{bi} \right)  \partial_{\mu_1\dots\mu_{m-1}\alpha_1\dots\alpha_{N-1}} \phi_i \nonumber\\
&&
+  
\binom{m}{2}
{\tilde K}^{(\beta\gamma\mu_1\dots\mu_{m-2})}_{b} \left( \partial_{\beta\gamma} {\bar A}^{\alpha_1\dots\alpha_{N-1}}_{bi} \right)  \partial_{\mu_1\dots\mu_{m-2}\alpha_1\dots\alpha_{N-1}} \phi_i
\nonumber \\
&&
={\tilde K}^{(\mu_1\dots\mu_m)}_{b}  {\cal A}^{\alpha_1\dots\alpha_{N-1}}_{2,ba} \partial_{\mu_1\dots\mu_m\alpha_1\dots\alpha_{N-1}} \psi_a 
\nonumber \\ &&
+\left[{\tilde K}^{(\mu_1\dots\mu_m)}_{b} {\cal B}^{\alpha_1\dots\alpha_{N-2}}_{2,ba} + m {\tilde K}^{(\beta \mu_1\dots\mu_{m-1})}_{b}   \left( \partial_\beta {\cal A}^{\mu_m \alpha_1\dots \alpha_{N-2} }_{2,ba}  \right)  \right] \partial_{\mu_1\dots\mu_m\alpha_1\dots\alpha_{N-2}} \psi_a
\nonumber \\ &&
+ \cO { \partial^{m+N-3} \psi } , 
\label{E2}
\end{eqnarray}
where
\begin{eqnarray}
&&{\cal A}^{\alpha_1\dots\alpha_{N-1}}_{2,ba}:=  {\bar A}^{\alpha_1\dots\alpha_{N-1}}_{bi} B_{ia} + (N-1)  {\bar A}^{\beta(\alpha_1\dots\alpha_{N-2}}_{bi}\left(\partial_\beta  A^{\alpha_{N-1})}_{ia}\right) ,\\
&&{\cal B}^{\alpha_1\dots\alpha_{N-2}}_{2,ba} := 
(N-1)\left[  {\bar A}^{\beta \alpha_1\dots\alpha_{N-2}}_{bi} \left(\partial_\beta B_{ia} \right) + \frac{N-2}{2}  {\bar A}^{\beta\gamma (\alpha_1\dots\alpha_{N-3}}_{bi}\left(\partial_{\beta\gamma}  A^{\alpha_{N-2})}_{ia}\right)  \right]
\end{eqnarray}
The right-hand side of Eq.~(\ref{E2}) has one less derivative compared to that of Eq.~\eq{eq:C}, that is, it is the subleading equation. 
The combination of non-degenerate part of Eq.~\eq{eq:C}, which has $N-1$ equations,
and Eq.~\eq{E2} gives the structure of the characteristics in the subleading order. 

So far we established the presence of one degeneracy, meanwhile to ensure invertibility, there must be $N$ degeneracies.
Hence
the characteristics at this subleading order should be degenerate too. 
By assumption, the matrix $A^\mu_{ia}\xi_\mu$ has one degeneracy, that is, there exist only one eigenvector with zero eigenvalue, 
which is denoted $\psi_a^\perp (\xi)$. 
The other components of $\psi_a$ are collectively defined as $\psi_a^\parallel(\xi)$. 
(Hereinafter,  we omit $(\xi)$ of  $\psi_a^\perp (\xi)$ and $\psi_a^\parallel(\xi)$ for brevity.) 
The characteristic matrix for the highest derivative part of Eq.~\eq{E2} and non-degenerate part of Eq.~\eq{eq:C} is written as
\begin{equation}
\bordermatrix{     & \psi^\parallel& \psi^\perp \cr
              \mbox{non-degenerate part of Eq.~(\ref{eq:C})} \hspace{5mm} &{}^{\rm(nd)} \! A^\mu_{ia} \xi_\mu &  0  \cr
	      \hfill
	      \mbox{Eq.~(\ref{E2})} \hspace{3mm} & K_{a} & K 
            } \ . \label{Mat}
\end{equation}
where $(K_a,K) ={\tilde K}^{(\mu_1\dots\mu_m)}_{b}  {\cal A}^{\alpha_1\dots\alpha_{N-1}}_{2,ba} \xi_{\mu_1} \dots \xi_{\mu_m} \xi_{\alpha_1}\dots \xi_{\alpha_{N-1}}$,\footnote{
Using the projection tensor $E_{N-1, ab}$ onto the $\psi^\parallel$ space defined by Eq.~(\ref{tilKtodbtilK}), the components of the vector $(K_a, K)$ may be expressed more precisely as 
\begin{align*}
K_a&\propto {\tilde K}^{(\mu_1\dots\mu_m)}_{b}  {\cal A}^{\alpha_1\dots\alpha_{N-1}}_{2,bc} E_{N-1,cb} \xi_{\mu_1} \dots \xi_{\mu_m} \xi_{\alpha_1}\dots \xi_{\alpha_{N-1}}~,\\
K&\propto {\tilde K}^{(\mu_1\dots\mu_m)}_{b}  {\cal A}^{\alpha_1\dots\alpha_{N-1}}_{2,bc} \bar A_{ci}^{\nu_1\dots\mu_{N-1}}\bar A_{ai}^{\nu_N\dots \nu_{2N-2}}\xi_{\mu_1} \dots \xi_{\mu_m} \xi_{\alpha_1}\dots \xi_{\alpha_{N-1}}\xi_{\nu_1} \dots \xi_{\nu_{2N-2}}~.
\end{align*}
See section~\ref{sec:subsubleading} for more details on this decomposition.
} and ${}^{\rm(nd)} \! A^\mu_{ia} \xi_\mu$ is a $(N-1)\times (N-1)$ matrix indicating only the non-degenerate components of $A^\mu_{ia} \xi_\mu$, that is, $\det \bigl({}^{\rm(nd)} \! A^\mu_{ia} \xi_\mu\bigr) \neq 0$. 
This matrix determines the subleading characteristics, and it should be degenerate for the invertibility. 
Since ${}^{\rm(nd)} \! A^\mu_{ia} \xi_\mu$ is regular, the requirement of the degeneracy gives the condition that $K=0$ for any ${\tilde K}^{(\mu_1\dots\mu_m)}_{b}$, {\it i.e.} $ {\cal A}^{\alpha_1\dots\alpha_{N-1}}_{2,ba} \xi_{\alpha_1}\dots \xi_{\alpha_{N-1}}\psi^\perp_a $ vanishes for any $\xi$.

In order to show the condition explicitly, we express $\psi^\perp$ in terms of $A^\mu_{ia}$.  
It can be done with the adjugate matrix ${\bar A}^{\mu_1\dots\mu_{N-1}}_{ai} \xi_{\mu_1} \dots\xi_{\mu_{N-1}}$. 
Equation \eq{AA} shows that ${\bar A}^{\mu_1\dots\mu_{N-1}}_{ai} \xi_{\mu_1} \dots\xi_{\mu_{N-1}}$ is the projection matrix 
to the kernel of $A^\mu_{ia} \xi_\mu$, 
in which $\psi^\perp$ lives by definition.
Therefore, the subleading degeneracy condition is written as
\begin{equation}
{\cal A}^{(\alpha_1\dots\alpha_{N-1}}_{2,ba} \bar A_{ai}^{\mu_1 \dots\mu_{N-1})}  =0.
\label{sub}
\end{equation}

\subsubsection{
Degeneracy and non-degeneracy conditions at lower orders}
\label{sec:subsubleading}

So far, two degeneracies were realized by imposing the degeneracy conditions (\ref{LO}) and (\ref{sub}).
The condition \eq{sub}
makes the matrix \eq{Mat} degenerate,
and then,  the highest order derivative of the component parallel to $\psi^\perp$ appears in a lower order equation, 
that is, the subsubleading order equation. 
The characteristic matrix is composed of the leading order equation for the $\psi^\parallel$ components and 
of the subsubleading order equation for the $\psi^\perp$ component. 
Now we have two degeneracy, and thus,
for $N>2$ this subsubleading characteristic matrix must give additional degeneracy for invertibility. 
Then, similar to the analysis of the subleading order, 
the degeneracy implies
that the highest order derivative for $\psi^\perp$ component 
in the subsubleading equation should vanish. 
This procedure is done iteratively until $N$ degeneracies are established. 
After that, we impose the condition that the next-order characteristic matrix is \emph{not} degenerate, in order for the  transformation to be invertible.
The last condition corresponds to that of the inverse function theorem (without derivatives). 

In order to implement the procedure described above, 
it is useful to introduce a projection matrix $E_{N-1,ij}(\xi)$ to the $(N-1)$-dimensional field space of $\psi^\parallel$, 
that is, $E_{N-1,ij}(\xi)$ is the identity matrix for the $(N-1)$-dimensional field space and zero for the one-dimensional field space parallel to $\psi^\perp$. 
Let us express $E_{N-1,ij}(\xi)$ in terms of
$A^\mu_{ia}\xi_\mu$.
We consider a matrix
\begin{equation}
\tilde A^{\mu_1\dots \mu_{2N-3}}_{ai}
=
\frac{1}{(N-1)!}
\epsilon_i{}^{i_1\dots i_{N-1}}
\epsilon_a{}^{a_1\dots a_{N-1}}
{\bar A}^{(\mu_1\dots \mu _{N-1}}_{a_1i_1} A^{\mu_N}_{i_2 a_2} \dots A^{\mu_{2N-3})}_{i_{N-1} a_{N-1}}  .
\end{equation}
This matrix satisfies
\begin{align}
&(N-1)\tilde A^{\mu_1\dots \mu_{2N-3}}_{ai} A^{\mu_{2N-2}}_{ib} \xi_{\mu_1}\dots \xi_{\mu_{2N-2}}
\nonumber \\& \qquad
=\left(
  {\bar A}^{2,\mu_1\ldots \mu_{2N-2}}  \delta_{ab}
- {\bar A}_{a i}^{\mu_1\ldots \mu_{N-1}} {\bar A}_{b i }^{\mu_N\ldots\mu_{2N-2}}
\right)
\xi_{\mu_1}\dots \xi_{\mu_{2N-2}}
\nonumber\\& \qquad
=
E_{N-1,ab} (\xi)
{\bar A}^{2,\mu_1\dots \mu _{2N-2}} \xi_{\mu_1}\dots \xi_{\mu_{2N-2}},
\label{calE}
\end{align}
where
\begin{equation}
{\bar A}^{2,\mu_1\dots \mu _{2N-2}} := {\bar A}_{ai}^{(\mu_1\dots \mu _{N-1}}{\bar A}_{ai}^{\mu_N\dots \mu _{2N-2})}
=\tilde A^{(\mu_1\dots \mu_{2N-3}}_{ai} A^{\mu_{2N-2})}_{ia} .
\label{barA2}
\end{equation}
In the calculation in Eq.~\eq{calE}, we use the fact that ${\bar A}_{a i}^{\mu_1\ldots \mu_{N-1}} \xi_{\mu_1}\dots \xi_{\mu_{N-1}}$ is rank-1 matrix, and then the components of 
${\bar A}_{a i}^{\mu_1\ldots \mu_{N-1}} {\bar A}_{b i }^{\mu_N\ldots\mu_{2N-2}}\xi_{\mu_1}\dots \xi_{\mu_{2N-2}}$ 
are zero except for 
the $\psi^\perp$-$\psi^\perp$ component.
The value of this nonzero component is shown to be 
${\bar A}^{2,\mu_1\dots \mu _{2N-2}}\xi_{\mu_1}\dots \xi_{\mu_{2N-2}}$
by direct calculations.
Equation \eq{calE} divided by ${\bar A}^{2,\mu_1\dots \mu _{2N-2}}\xi_{\mu_1}\dots \xi_{\mu_{2N-2}}$ gives 
$E_{N-1,ab} (\xi)$, 
which can be regarded as a projector
to the $(N-1)$-dimensional space of $\psi^\parallel$.
Since Eq.~\eq{calE} holds for any $\xi^\mu$, it can be expressed equivalently as
\begin{equation}
 {\bar A}_{ai}^{(\mu_1\ldots \mu_{N-1}} {\bar A}_{bi}^{\mu_N\ldots\mu_{2N-2)}}+(N-1)\tilde A^{(\mu_1\dots \mu_{2N-3}}_{ai} A^{\mu_{2N-2)}}_{ib} 
=  {\bar A}^{2,(\mu_1\ldots \mu_{2N-2})}  \delta_{ab} .
\label{project}
\end{equation}

Let us demonstrate how to obtain the lower order equations iteratively with the projection tensor~\eq{calE}.
On the right-hand side of Eq.~(\ref{E2}),  the leading term is the first term
proportional to $\partial^{m+N-1}\psi_a$.
The degeneracy condition for the subleading order \eq{sub} implies that the first term does not have the 
$\psi^\perp$ component, {\it i.e.} it can be written as
\begin{equation}
{\tilde K}^{(\mu_1\dots\mu_m)}_{b}  {\cal A}^{\alpha_1\dots\alpha_{N-1}}_{2,ba} \partial_{\mu_1\dots\mu_m\alpha_1\dots\alpha_{N-1}} \psi_a = {\tilde K}^{(\mu_1\dots\mu_m)}_{b}  {\cal A}^{\alpha_1\dots\alpha_{N-1}}_{2,bc} E_{N-1,cd}(\partial) \partial_{\mu_1\dots\mu_m\alpha_1\dots\alpha_{N-1}} \psi_d.
\label{tilKtodbtilK}
\end{equation}
Considering the case with ${\tilde K}^{\mu_1\dots\mu_m}_{b}= {\dbtilde K}^{\mu_1\dots\mu_{m'}}_{b} {\bar A}^{2,\gamma_1\dots\gamma_{2N-2}}$,
we can rewrite Eq.~\eq{tilKtodbtilK} as
\begin{eqnarray}
&&{\dbtilde K}^{\mu_1\dots\mu_{m'}}_{b} {\bar A}^{2,\gamma_1\dots\gamma_{2N-2}}  {\cal A}^{\alpha_1\dots\alpha_{N-1}}_{2,bc} E_{N-1,ca}(\partial) \partial_{\mu_1\dots\mu_{m'}\gamma_1\dots\gamma_{2N-2}\alpha_1\dots\alpha_{N-1}} \psi_k 
\nonumber \\
&&\qquad
=
(N-1)
{\dbtilde K}^{\mu_1\dots\mu_{m'}}_{b}  {\cal A}^{\alpha_1\dots\alpha_{N-1}}_{2,bc}  \tilde A^{\gamma_1\dots \gamma_{2N-3}}_{ci} A^{\gamma_{2N-2}}_{ia} \partial_{\mu_1\dots\mu_{m'}\alpha_1\dots\alpha_{N-1}\gamma_1\dots\gamma_{2N-2}} \psi_k 
\end{eqnarray}
where we used Eq.~(\ref{calE}).
Then, by subtracting Eq.~\eq{E2} from Eq.~(\ref{eq:C}) with its coefficient set to 
$K^{\mu_1\dots\mu_m}_{bi} =(N-1)
{\dbtilde K}^{\mu_1\dots\mu_{m'}}_{b}  {\cal A}^{\alpha_1\dots\alpha_{N-1}}_{2,bc}  \tilde A^{\gamma_1\dots \gamma_{2N-3}}_{ci}$,
the subleading ($\partial^{m'+3N-3}\psi$) term is canceled out and the subsubleading-order equation is obtained as a result.

The degeneracy condition can be obtained applying the procedure explained around \eq{Mat} to the subsubleading equation, and the condition will be similar to the condition \eq{sub} for the subleading-order equation.
Then, using the projector $E_{N-1,ab}$ again, we can construct the lower-order equation.
Applying this procedure iteratively, we can derive $N$ degeneracy conditions and the final non-degeneracy condition, 
which constitute the necessary conditions for the invertibility.

\subsection{Necessary conditions in two-field case}
\label{sec:two-fields}

Although we have already derived the leading and subleading degeneracy conditions in the previous subsection 
using the general procedure, it is 
instructive to follow 
a concrete example for understanding the method.
For this purpose,
based on the discussion in the previous section, 
we demonstrate how to derive the necessary conditions for the invertibility 
of a field transformation 
of two fields 
involving up to first derivative.\footnote{As will be explicitly shown in Appendix \ref{App:one-field}, there is no invertible field transformation of one field involving first derivatives.
Hence, among (possibly invertible) transformations with up to first derivatives, the two field case is the simplest.} 
That is,
we consider the case where $N=2$, $i=1,2$ and $a=1,2$. 
In this section,
we will complete the iterations to derive all the degeneracy and the final non-degeneracy conditions.
The obtained conditions will be simplified in section \ref{sec:2sim}.

We apply the procedure explained in section~\ref{sec:necessary_general} 
to the two-field case in this section.
In this case, ${\bar A}^\mu_{ai}$, ${\cal A}^\mu_{2,ab}$ and ${\cal B}_{2,ab}$ appeared in the previous section are written as
\begin{equation}
{\bar A}^\mu_{ai}=\epsilon_i{}^{i_1}\epsilon_a{}^{a_1}A^\mu_{i_1 a_1}, \qquad
{\cal A}^\mu_{2,ab}:= \bar A^\mu_{ai} B_{ib} + \bar A^\beta_{ai} \partial_\beta  A^{\mu}_{ib},
\qquad
{\cal B}_{2,ab} :=   {\bar A}^{\beta }_{ai} \partial_\beta B_{ib}~,
\label{calAdef}
\end{equation}
and ${\tilde A}^\mu_{ai}$ is simplified as
\begin{equation}
{\tilde A}^\mu_{ai} 
= \epsilon_i{}^{i_1}\epsilon_a{}^{a_1} {\bar A}^\mu_{a_1 i_1} 
= \epsilon_i{}^{i_1}\epsilon_a{}^{a_1}\epsilon_{i_1}{}^{i_2}\epsilon_{a_1}{}^{a_2}A^\mu_{i_2a_2}
= A^\mu_{ia}.
\end{equation}
Then the identity (\ref{project}) becomes
\begin{equation}
 {\bar A}_{ai}^{(\mu_1} {\bar A}_{bi}^{\mu_2)}+\tilde A^{(\mu_1}_{ai} A^{\mu_2)}_{ib}
 =
  {\bar A}_{ai}^{(\mu_1} {\bar A}_{bi}^{\mu_2)} + A^{(\mu_1}_{ia} A^{\mu_2)}_{ib}
=  {\bar A}^{2,\mu_1\mu_2}  \delta_{ab} .
\label{project_2field}
\end{equation}

We operate $K^{(\mu_1\dots\mu_n)}\partial_{\mu_1\dots \mu_n}$ on the field transformation equation \eq{redef} to obtain
\begin{equation}
K^{(\mu_1\dots\mu_n)}\partial_{\mu_1\dots \mu_n} \phi_i
=K^{(\mu_1\dots\mu_n)}A^{\mu_{n+1}}_{ia}\partial_{\mu_1\dots \mu_{n+1}} \psi_a   + \cO{\partial^{n} \psi}. 
\label{leading_2fields}
\end{equation}
For \eq{redef} to be invertible, the coefficient of the highest-order derivative on the right-hand side Eq.~(\ref{leading_2fields}) must be degenerate, which implies
\begin{equation}
{}^\forall \xi_\mu\,, \quad 
\det (A^\mu_{ia} \xi_\mu)=0 \qquad 
\Leftrightarrow
\qquad  \epsilon^{i_1 i_2}\epsilon^{a_1 a_2} A^{(\alpha_1}_{i_1a_1} A^{\alpha_2)}_{i_2 a_2}=0.
\label{detA}
\end{equation}
Following the previous section, we assume that the matrix $A^\mu_{ia} \xi_\mu$ is degenerate only in one dimension. Then the kernel of $A^\mu_{ia} \xi_\mu$ parallel to $\psi_a^\perp$ and the field space in the other direction parallel to $\psi_a^{\smash \parallel}$ are one dimension each.

The degeneracy condition at the subleading order is obtained
following the procedure in section~\ref{sec:subleading}.
To construct
the subleading equation from Eq.~\eq{redef}, we have to pick up the component of this equation corresponding to the kernel of the coefficient of the highest order derivative, $A^\mu_{ia}$. 
It can be done by operating $\tilde K^{(\mu_1\dots\mu_n)} \bar A^{\mu_{n+1}}_{bi}\partial_{\mu_1\dots \mu_{n+1}} $ to the transformation equation \eq{redef} as
\begin{align}
&\tilde K^{(\mu_1\dots\mu_n)} \bar A^{\mu_{n+1}}_{bi}\partial_{\mu_1\dots \mu_{n+1}} \phi_i
\nonumber \\
&=\tilde K^{(\mu_1\dots\mu_n)} \bar A^{\mu_{n+1}}_{bi}A^{\nu}_{ia}\partial_{\mu_1\dots \mu_{n+1}\nu} \psi_a
\nonumber \\
&
+ \left( \tilde K^{(\mu_1\dots\mu_n)} \bar A^{\mu_{n+1}}_{bi} B_{ia}
+ n \, \tilde K^{(\alpha\mu_1\dots\mu_{n-1})} \bar A^{\mu_{n}}_{bi} \partial_\alpha A^{\mu_{n+1}}_{ia}
+\tilde K^{(\mu_1\dots\mu_n)} \bar A^{\alpha}_{bi} \partial_\alpha A^{\mu_{n+1}}_{ia}
\right)
\partial_{\mu_1\dots \mu_{n+1}} \psi_a 
\nonumber \\
&
+ \cO{\partial^{n} \psi}. 
\label{tilK2}
\end{align}
The first term on the right-hand side vanishes thanks to Eq.~\eq{detA}.
As shown in section~\ref{sec:subleading}, the second term in the brackets proportional to $n$ can be canceled by adding
$n{\tilde K}^{(\alpha\mu_1\dots\mu_{n-1})} \left( \partial_\alpha {\bar A}^{\mu_n}_{bi} \right)  \partial_{\mu_1\dots\mu_n} \phi_i $ to Eq.~(\ref{tilK2}).
Nevertheless, without such a cancellation, we can directly obtain the same subleading condition from Eq.~\eq{tilK2} by rewriting it as
\begin{eqnarray}
&&\tilde K^{(\mu_1\dots\mu_n)} \bar A^{\mu_{n+1}}_{bi}\partial_{\mu_1\dots \mu_{n+1}} \phi_i
\nonumber \\ 
&& \qquad
= 
\left( \tilde K^{(\mu_1\dots\mu_n)} {\cal A}^{\mu_{n+1}}_{2,ba} + n \, \tilde K^{(\alpha\mu_1\dots\mu_{n-1})} \bar A^{\mu_{n}}_{bi} \partial_\alpha A^{\mu_{n+1}}_{ia}
\right)
\partial_{\mu_1\dots \mu_{n+1}} \psi_a + \cO{\partial^{n} \psi}. 
\end{eqnarray}
The subleading degeneracy condition is given 
by demanding that the coefficient of highest-order derivative for $\psi^\perp$ vanishes. 
Noting that $\bar A^\nu_{ai} \xi_\nu$ works as a projector onto the $\psi^\perp$ space,
this condition is written as
\begin{equation}
\left( \tilde K^{(\mu_1\dots\mu_n)} {\cal A}^{\mu_{n+1}}_{2,ba} + n \, \tilde K^{(\alpha\mu_1\dots\mu_{n-1})} \bar A^{\mu_{n}}_{bi} \partial_\alpha A^{\mu_{n+1}}_{ia}
\right) \bar A^\nu_{aj} \xi_{\mu_1} \dots \xi_{\mu_{n+1}} \xi_\nu =0.
\end{equation}
The second term vanishes, which is shown by using Eq.~(\ref{AA}) and (\ref{AbardA=0}) as 
\begin{align}
&\tilde K^{(\alpha\mu_1\dots\mu_{n-1})} \bar A^{\mu_{n}}_{bi} \left(\partial_\alpha A^{\mu_{n+1}}_{ia}\right)\bar A^\nu_{aj} \xi_{\mu_1} \dots \xi_{\mu_{n+1}} \xi_\nu 
\notag \\ &\quad
=
-
\tilde K^{(\alpha\mu_1\dots\mu_{n-1})} \left(\partial_\alpha \bar A^{\mu_{n}}_{bi} \right) A^{\mu_{n+1}}_{ia}\bar A^\nu_{aj} \xi_{\mu_1} \dots \xi_{\mu_{n+1}} \xi_\nu =0.    
\end{align}
Since $\tilde K^{(\mu_1\dots\mu_n)}$ is arbitrary, the subleading condition is 
\begin{equation}
{}^\forall \xi_\mu\,, \quad
 {\cal A}^{\mu}_{2,ba} \bar A^\nu_{aj} \xi_\mu \xi_\nu =0 \qquad
\Leftrightarrow
\qquad 
 {\cal A}^{(\mu}_{2,ba} \bar A^{\nu)}_{aj}
 =
\bar A^{(\mu}_{bi} B_{ia} \bar A^{\nu)}_{aj} + \bar A^\beta_{bi}\left( \partial_\beta  A^{(\mu}_{ia}\right)\bar A^{\nu)}_{aj}=0 .
\label{2sub}
\end{equation}

The non-degeneracy condition at the subsubleading order can be constructed following the procedure in section~\ref{sec:subsubleading}.
In this procedure 
we need to see the structure of the coefficient of highest-order derivative term in Eq.~\eq{tilK2} 
decomposing the variables into the $\psi^\perp$ and $\psi^\parallel$ space. For this purpose we use Eq.~\eq{project_2field}.
Operating $\dbtilde K^{(\mu_1\dots\mu_n)} \bar A^{\mu_{n+1}}_{bi} \bar A^{2,\mu_{n+2}\mu_{n+3}} \partial_{\mu_1\dots \mu_{n+3}} $ to Eq.~\eq{redef}, we have
\begin{eqnarray}
&&\dbtilde K^{(\mu_1\dots\mu_n)} \bar A^{\mu_{n+1}}_{bi} \bar A^{2,\mu_{n+2}\mu_{n+3}} \partial_{\mu_1\dots \mu_{n+3}} \phi_i
\nonumber \\ 
&& \qquad
= \left( \dbtilde K^{(\mu_1\dots\mu_n)} {\cal A}^{\mu_{n+1}}_{2,ba} \bar A^{2,\mu_{n+2}\mu_{n+3}} + n\dbtilde K^{(\alpha\mu_1\dots\mu_{n-1})} \bar A^{2,\mu_{n}\mu_{n+1}} \bar A^{\mu_{n+2}}_{bi} \partial_\alpha A^{\mu_{n+3}}_{ia}
\right.
\nonumber \\ 
&& \left.\hspace{25mm}
+2 \dbtilde K^{(\mu_1\dots\mu_{n})} \bar A^{2,\mu_{n+1}\alpha} \bar A^{\mu_{n+2}}_{bi} \partial_\alpha A^{\mu_{n+3}}_{ia}
\right)
\partial_{\mu_1\dots \mu_{n+3}} \psi_a + \cO{\partial^{n+2} \psi}. 
\end{eqnarray}
Using Eq.~\eq{project_2field} for the first term in brackets and
Eq.~(\ref{AbardA=0})
for the other two, 
we have 
\begin{eqnarray}
&&\dbtilde K^{(\mu_1\dots\mu_n)} \bar A^{\mu_{n+1}}_{bi} \bar A^{2,\mu_{n+2}\mu_{n+3}} \partial_{\mu_1\dots \mu_{n+3}} \phi_i
\nonumber \\ 
&& \qquad
= \left( \dbtilde K^{(\mu_1\dots\mu_n)} {\cal A}^{\mu_{n+1}}_{2,bc} \tilde A^{\mu_{n+2}}_{ci} - n\dbtilde K^{(\alpha\mu_1\dots\mu_{n-1})} \bar A^{2,\mu_{n}\mu_{n+1}}
\partial_\alpha \bar A^{\mu_{n+2}}_{bi}
\right.
\nonumber \\ 
&& \left.\hspace{25mm}
-2 \dbtilde K^{(\mu_1\dots\mu_{n})} \bar A^{2,\mu_{n+1}\alpha}
\partial_\alpha \bar A^{\mu_{n+2}}_{bi}
\right) A^{\mu_{n+3}}_{ia}
\partial_{\mu_1\dots \mu_{n+3}} \psi_a + \cO{\partial^{n+2} \psi}. 
\label{subtilK2}
\end{eqnarray}
The $\partial^{n+3}\psi$ term
in the right-hand side can be reproduced by applying the operator
\begin{equation*}
\left[
\dbtilde K^{(\mu_1\dots\mu_n)} {\cal A}^{\mu_{n+1}}_{2,bc} \tilde A^{\mu_{n+2}}_{ci}
-
\left(
n\dbtilde K^{(\alpha\mu_1\dots\mu_{n-1})}\bar A^{2,\mu_{n}\mu_{n+1}}
+2 \dbtilde K^{(\mu_1\dots\mu_{n})} \bar A^{2,\mu_{n+1}\alpha}
\right)
\partial_\alpha \bar A^{\mu_{n+2}}_{bi} 
\right]
\partial_{\mu_1\dots \mu_{n+2}}
\end{equation*}
to  Eq.~\eq{redef}. 
Hence, by subtracting it from Eq.~\eq{subtilK2}, we obtain the subsubleading-order equation involving terms only up to $\partial^{n+2}\psi$.
Expanding Eqs.~\eq{eq:C} and Eq.~\eq{subtilK2} up to the $\partial^{n+2}\psi$ terms, we find (see Appendix~\ref{App:longcalculation} for details)
\begin{eqnarray}
 &&\dbtilde K^{(\mu_1\dots\mu_n)} \bar A^{\mu_{n+1}}_{bi} \bar A^{2,\mu_{n+2}\mu_{n+3}} \partial_{\mu_1\dots \mu_{n+3}} \phi_i 
\nonumber \\ &&
-
\left[
\dbtilde K^{(\mu_1\dots\mu_n)} {\cal A}^{\mu_{n+1}}_{2,bc} \tilde A^{\mu_{n+2}}_{ci}
-
\left(
n\dbtilde K^{(\alpha\mu_1\dots\mu_{n-1})}\bar A^{2,\mu_{n}\mu_{n+1}}
+2 \dbtilde K^{(\mu_1\dots\mu_{n})} \bar A^{2,\mu_{n+1}\alpha}
\right)
\partial_\alpha \bar A^{\mu_{n+2}}_{bi} 
\right]
\partial_{\mu_1\dots \mu_{n+2}} \phi_i
\nonumber \\ 
&&= \bigg\{ 
-\frac{n(n-1)}{2} \dbtilde K^{(\alpha_1\alpha_2\mu_1\dots\mu_{n-2})}  \bar A^{2,\mu_{n-1}\mu_n} 
\left( \partial _{\alpha_1 \alpha_2} \bar A^{\mu_{n+1}}_{bi} \right) A^{\mu_{n+2}}_{ia}  
\nonumber \\ && \quad
+n\,\dbtilde K^{(\alpha_1 \mu_1\dots\mu_{n-1})} \left[
-2\bar A^{2,\mu_n\alpha_2} 
\left(\partial_{\alpha_1\alpha_2} \bar A^{\mu_{n+1}}_{bi}\right)
A^{\mu_{n+2}}_{ia}
 -{\cal A}^{\mu_{n}}_{2,bc} \tilde A^{\mu_{n+1}}_{ci}  \partial_{\alpha_1} A^{\mu_{n+2}}_{ia} 
+\bar A^{2,\mu_{n}\mu_{n+1}} \partial_{\alpha_1} {\cal A}^{\mu_{n+2}}_{ba} 
\right]
\nonumber \\ && \quad
+\dbtilde K^{(\mu_1\dots\mu_{n})}
\Bigl[
- \bar A^{2,\alpha_1\alpha_2} \left( \partial_{\alpha_1\alpha_2}\bar A^{\mu_{n+1}}_{bi} \right)  A^{\mu_{n+2}}_{ia} +2\bar A^{2,\mu_{n+1}\alpha} \partial_{\alpha} {\cal A}^{\mu_{n+2}}_{2,ba}
+ \bar A^{2,\mu_{n+1}\mu_{n+2}}  \bar A^{\alpha}_{bi}\partial_{\alpha} B_{ia}
\nonumber \\ && \hspace{24mm}
 -2 {\cal A}^{(\alpha}_{2,bc} \tilde A^{\mu_{n+1})}_{ci}
\partial_{\alpha} A^{\mu_{n+2}}_{ia} 
-{\cal A}^{\mu_{n+1}}_{2,bc} \tilde A^{\mu_{n+2}}_{ci}B_{ia}
\Bigr]
\biggr\} \partial_{\mu_1\dots \mu_{n+2}} \psi_a
+\cO{\partial^{n+1} \psi_a}~.
\label{subsubleadingeq_2field}
\end{eqnarray}
The coefficient of $\psi^\perp$ component of the highest order derivative term $\partial^{n+2}\psi$ on the right-hand side of Eq.~(\ref{subsubleadingeq_2field}) determines the (non)degeneracy of the characteristics in the subsubleading order. 
This coefficient
is obtained by contracting $\bar A^{\mu_{n+3}}_{ai}$ with the coefficient of $\partial^{n+2}\psi_a$ in Eq.~(\ref{subsubleadingeq_2field}). 
We can show that the terms proportional to $\dbtilde K^{(\alpha_1 \alpha_2\mu_1\dots\mu_{n-2})}$ and  $\dbtilde K^{(\alpha_1 \mu_1\dots\mu_{n-1})}$ become zero (see Appendix~\ref{App:longcalculation}).
Among the terms proportional to $\dbtilde K^{(\mu_1\dots\mu_n)}$, the first one ($\partial_{\alpha_1\alpha_2}\bar A^{\mu_{n+1}}_{bi}$ term) vanishes by acting $\bar A^{\mu_{n+3}}_{ai}$, and the remaining term must not vanish for the $\psi^\perp$ component to be non-degenerate in this order.
Hence, the non-degeneracy condition is given by 
\begin{equation}
\left( \bar A^{2,\mu_1\mu_2} {\cal B}_{2,ab}- {\cal A}^{\mu_1}_{2,ac} \tilde A^{\mu_2}_{cj}   B_{jb}+ 2\bar A^{2,\alpha\mu_1} \partial_\alpha {\cal A}^{\mu_2}_{2,ab}
- 2{\cal A}^{(\alpha}_{2,ac} \tilde A^{\mu_1)}_{cj} \partial_\alpha A^{\mu_2}_{jb}
\right)
\bar A^{\mu_3}_{bi} \xi_{\mu_1}\xi_{\mu_2}\xi_{\mu_3} \neq 0
\label{2sspre}
\end{equation}
for any $\xi$.
Using Eq.~(\ref{project_2field}), this can be also written as  
\begin{equation}
\left( \bar A^{2,\mu_1\mu_2} {\cal B}_{2,ab}\bar A^{\mu_3}_{bi}
- {\cal A}^{\mu_1}_{2,ab} \tilde A^{\mu_2}_{bj}   B_{jc}\bar A^{\mu_3}_{ci}
+2{\cal A}_{2,ab}^{[\alpha|} {\tilde A}_{bj}^{\mu_1} A_{jc}^{|\mu_2]} \partial_\alpha {\bar A}_{ci}^{\mu_3}
-{\cal A}_{2,ab}^{\mu_1} {\bar A}_{bj}^{\alpha} {\bar A}_{cj}^{\mu_2} \partial_\alpha {\bar A}_{ci}^{\mu_3}
\right)
\xi_{\mu_1}\xi_{\mu_2}\xi_{\mu_3}
 \neq 0.
\label{2ss}
\end{equation}

To summarize, the necessary conditions for the invertibility are given by Eqs.~(\ref{detA}), (\ref{2sub}) and (\ref{2ss}).

\subsection{Simplification of necessary conditions}\label{sec:2sim}

We have derived above the necessary conditions for the invertibility (\ref{detA}), (\ref{2sub}) and (\ref{2ss}) in the two-field case with first-order derivative. 
We present a simpler form of these expressions in this section. 
The simplified form of the necessary conditions will be helpful in the proof that these conditions are also sufficient conditions shown in the next section. In addition, we will use these expressions for building examples of invertible transformations.

\subsubsection{Simplification of condition (\ref{detA})}
\label{sec:2sim_1}

First of all, we simplify  condition (\ref{detA}). 
Equation (\ref{detA}) implies that the rank of $A_{ia}^\mu \xi_\mu$ is less than two. 
Since we have assumed that $A_{ia}^\mu \xi_\mu$ has only one eigenvector with zero eigenvalue, 
the rank of $A_{ia}^\mu \xi_\mu$ has to be unity for any $\xi$, that is,
\begin{equation}
\xi_0 A_{ia}^0+\xi_1 A_{ia}^1+\dots+ \xi_D A_{ia}^D \label{xiA}
\end{equation}
is a rank-one matrix for any $\xi_\mu =\{ \xi_0,\xi_1,\dots,\xi_D \}$, 
where $D$ is the spacetime dimension.
Then, by setting $\xi_\mu = \delta_{\mu\nu}$ for each $1\leq \nu \leq D$, 
we find that $A^\nu_{ia}$ is a rank-one matrix for any $\nu$.
Any rank-one 2$\times$2 matrix can be written by a product of vectors, {\it i.e.} the matrices 
${A_{ia}^0, A_{ia}^1, \dots, A_{ia}^D}$ are written as
\begin{equation}
A_{ia}^0 = V_i^0 U_a^0,\qquad A_{ia}^1 = V_i^1 U_a^1,\qquad \dots \qquad A_{ia}^D = V_i^D U_a^D.  
\end{equation}

Below, we show by the induction that, if the rank of (\ref{xiA}) is one for any $\xi$, 
$A_{ia}^\mu$ can be written as
\begin{equation}
A_{ia}^\mu = V_i U_a^\mu\qquad \mbox{or} \qquad A_{ia}^\mu = V_i^\mu U_a~.  \label{Aind}
\end{equation}
For $\mu\le 0$ (that is, $\mu=0$), $A_{ia}^0$ is written as $V_i^0 U_a^0$, and thus, regarding $(V_i^0, U_a^0)$ as  $(V_i, U_a^\mu)$ or $(V_i^\mu, U_a)$ for $\mu=0$, Eq.~\eq{Aind} is satisfied.  
Then, what we need to show is that, for any integer $k$, Eq.~\eq{Aind} satisfied for $\mu\le k+1$ if it is satisfied for $\mu\le k$.
Since
two choices of (\ref{Aind}) are symmetric with respect to $U_a$ and $V_i$, 
without loss of generality we assume the former ($A_{ia}^\mu = V_i U_a^\mu$) is satisfied for $\mu\le k$.
If $A_{ia}^{k+1}=0$, $A_{ia}^\mu = V_i U_a^\mu$ for $\mu\le k+1$ is trivially satisfied.
Therefore, we consider the case where $A_{ia}^{k+1} = V_i^{k+1}U_a^{k+1}\neq0$.
If $V_i^{k+1}$ is parallel to $V_i$,
by rescaling $V_i^{k+1}\to c V_i^{k+1}=V_i$, $U_a^{k+1}\to U_a^{k+1}/c$ with a constant $c$, 
we can satisfy $A_{ia}^\mu = V_i U_a^\mu$ for $\mu\le k+1$.

We show that if $V_i^{k+1}$ is not parallel to $V_i$, all $U_a^\mu$'s for $\mu \le k+1$ become parallel, and thus 
$A_{ia}^\mu$ can be written as $V_i^\mu U_a$.
We consider a vector $\xi_\mu=(\xi_0,\xi_1,\dots,\xi_{k+1},0,0,\dots)$, where $\xi_\mu$'s ($\mu \le k+1$) are arbitrary. 
Since $A_{ia}^\mu \xi_\mu$ should have an eigenvector $e^a(\xi_\mu)$ with zero eigenvalue, we have
\begin{equation}
0=A_{ia}^\mu \xi_\mu e^a = V_i \sum_{\mu=0}^k U_a^\mu e^a  \xi_\mu  + V^{k+1}_i U^{k+1}_a e^a  \xi_{k+1}~.
\end{equation}
If $V^{k+1}_i$ is not parallel to $V_i$, the above equation gives 
\begin{equation}
 \sum_{\mu=0}^k U_a^\mu \xi_\mu e^a   =0 ~, 
 \qquad
 U^{k+1}_a \xi_{k+1} e^a =0~.
 \label{Uxie}
\end{equation}
Now suppose that $\xi_{k+1}\neq0$. In this case,
the latter of these equations uniquely fixes $e^a$  
because the field space dimension is two.
Then, because $e^a$ is independent of $\xi_\mu$ ($\mu\leq k$)
and also $\xi^\mu$'s for $\mu \le k$ are arbitrary, 
Eq.~\eqref{Uxie} implies 
\begin{equation}
U_a^\mu e^a =0 \qquad  \bigl(\mu\le k+1\bigr)~.
\end{equation}
This implies that all $U_a^\mu$'s ($\mu\le k+1$) are normal to $e^a$. 
Since $e^a$ is non-zero and the space of index $a$ is two-dimensional, all $U_a^\mu$'s ($\mu\le k+1$) are parallel to each other.
Therefore, by rescaling $U_a$, $A_{ia}^\mu$ can be expressed in terms of the common $U_a$ as 
\begin{equation}
A_{ia}^\mu = C^\mu V_i U_a \quad (\mu\le k)~, \qquad  A_{ia}^{k+1} = V_i^{k+1} U_a,
\end{equation}
where $C^\mu$ is a normalization factor. 
Therefore, by redefining $\{C^\mu V_i^\mu, V_i^{k+1} \}$ $(\mu\le k)$ as $\{ V_i, V_i^{k+1}\}$, 
$A_{ia}^\mu$ is written as $V_i^\mu U_a$ for $\mu\le k+1$.

\subsubsection{Further decomposition of $A_{ia}^\mu$}
\label{sec:2sim_2}

We have seen that one of the invertibility conditions, Eq.~(\ref{detA}), implies that  $A_{ia}^\mu$ is decomposed as Eq.~\eq{Aind}. 
Using the other conditions, we show that $A_{ia}^\mu$ can be decomposed further as
\begin{equation}
A_{ia}^\mu = a^\mu V_i U_a. \label{aVU}
\end{equation}
Let us show the proof in each case of Eq.~\eq{Aind}.

\vspace{2mm}
{\noindent \underline{ i). $A_{ia}^\mu = V_i U_a^\mu$}} 
\vspace{2mm}

In this case, if we can show $U_a^\mu = a^\mu U_a$, $A_{ia}^\mu$ is expressed as $a^\mu V_i U_a$. 
We define a vector normal $n_i$ to $V_j$ as 
\begin{equation}
n_j:=\epsilon_{ij} V_j.
\end{equation} 
We also define a vector $m_a^\mu$ with spacetime index $\mu$ as  
\begin{equation}
m_a^\mu := \epsilon_{ab} U_b^\mu. 
\end{equation}
Then, $\bar A_{ia}^\mu$ and $\tilde A_{ia}^\mu$ are written as
\begin{equation}
\bar A_{ai}^\mu = m_a^\mu n_i~, \qquad
\tilde A_{ai}= U_a^\mu V_i~, 
\end{equation}
and condition (\ref{2sub}) becomes
\begin{equation}
n_i B_{ia} m_a^\mu =0 \qquad (\mbox{for any}\  \mu). 
\label{nBmmu}
\end{equation}

Now, we show that $U_a^\mu$ has to be written as $a^\mu U_a$. 
It can be proven by contradiction; 
the assumption  that two of $U_a^\mu$'s are not parallel to each other (for instance, $U_a^0$ is not parallel to $U_a^1$)
leads to a contradiction. 
Under this assumption, two of $m_a^\mu$'s are not parallel correspondingly.
Since the space spanned by index ``$a$'' is two-dimensional, 
Eq.~\eq{nBmmu} implies that 
\begin{equation}
 n_i B_{ia} = 0~.
 \label{nB=0}
\end{equation}
Then, the left hand side of (\ref{2ss}) becomes
\begin{eqnarray}
&&\biggl[
V_j^2 U_c^{\mu_1}U_c^{\mu_2} m_a^\beta n_k \left( \partial_\beta B_{kb}\right)m^{\mu_3}_b n_i 
-m^\beta_a U^{\mu_1}_b n_k  \left(\partial_\beta V_k \right) U^{\mu_2}_b V_j B_{jc} m^{\mu_3}_c n_i
\nonumber \\
&& \qquad
-m^\beta_a U^{\mu_1}_b n_k  \left(\partial_\beta V_k \right) m^{\alpha}_b n_j  m^{\mu_2}_c n_j  \partial_\alpha \left(  m^{\mu_3}_c n_i \right)
+m^\beta_a U^{\alpha}_b n_k  \left(\partial_\beta V_k \right) U^{\mu_1}_b V_j U^{\mu_2}_c V_j \partial_\alpha \left(  m^{\mu_3}_c n_i \right)
\nonumber \\
&& \qquad
-m^\beta_a U^{\mu_1}_b n_k  \left(\partial_\beta V_k \right) U^{\mu_2}_b V_j U^{\alpha}_c V_j \partial_\alpha \left(  m^{\mu_3}_c n_i \right)
\biggr] \xi_{\mu_1}\xi_{\mu_2}\xi_{\mu_3}
\nonumber \\
&& 
= \biggl\{
V_j^2 U_c^{\mu_1}U_c^{\mu_2} m_a^\beta m^{\mu_3}_b n_i 
\left[ n_k \left( \partial_\beta B_{kb}\right) + \left( \partial_\beta n_k  \right) B_{kb}\right] 
\nonumber \\
&& \qquad
- V_j^2 m^\beta_a  n_k  \left(\partial_\beta V_k \right) U^{\mu_1}_b U^{\mu_2}_b 
\left(  U^{\mu_3}_c    m^{\alpha}_c+U^{\alpha}_c    m^{\mu_3}_c\right) \left( \partial_\alpha n_i \right)
\nonumber \\
&& \qquad
- V_j^2 m^\beta_a  n_k  \left(\partial_\beta V_k \right)
\left(
 U^{\mu_1}_b m^{\alpha}_b  m^{\mu_2}_c  
-U^\alpha_b U^{\mu_1}_b U^{\mu_2}_c 
+ U^{\mu_1}_b U^{\mu_2}_b U^\alpha_c
\right)  \left( \partial_\alpha  m^{\mu_3}_c \right) n_i 
\biggr\} \xi_{\mu_1}\xi_{\mu_2}\xi_{\mu_3}
\nonumber \\
&& 
=0. \label{cont}
\end{eqnarray}
Here we used Eq.~\eqref{nB=0} and
\begin{equation}
\begin{aligned}
&{\cal A}^\mu_{2,ab} = m^\beta_a U^{\mu}_b n_k  \left(\partial_\beta V_k \right),\qquad
{\cal B}_{2,ab} = m_a^\beta n_i \partial _\beta B_{ib}, \qquad
\bar A^{2,\mu_1 \mu_2} = U_a^{\mu_1} U_a^{\mu_2} V_iV_i,
\\
&
m_a^\mu m_a^\nu =U_a^\mu U_a^\nu , \quad
U^{(\mu_1}_c    m^{\mu_2)}_c=0, \quad
V_k V_j +n_k n_j = V_i^2\delta_{kj}, \quad
m^{(\mu_1}_b  m^{\mu_2)}_c  + U^{(\mu_1}_b U^{\mu_2)}_c =U^{\mu_1}_a U^{\mu_2}_a \delta_{bc},
\\
&n_k \left( \partial_\beta V_k\right) V_j B_{jc}= - \left( \partial_\beta n_k\right)V_k V_j B_{jc} = - \left( \partial_\beta n_k\right)\left(V_k V_j +n_k n_j\right) B_{jc} =-V_k^2  \left( \partial_\beta n_j\right)B_{jc}, 
\\
&n_k \left( \partial_\beta B_{kb}\right) + \left( \partial_\beta n_k  \right) B_{kb} =\partial_\beta \left( n_k   B_{kb}\right) =0, 
\\
&  U^{(\mu_1|}_b m^{\alpha}_b  m^{|\mu_2)}_c  -U^\alpha_b U^{(\mu_1}_b U^{\mu_2)}_c + U^{(\mu_1}_b U^{\mu_2)}_b U^\alpha_c
= 
-U^{\alpha}_b \left(m^{(\mu_1}_b  m^{\mu_2)}_c + U^{\mu_1}_b U^{\mu_2}_c\right) + U^{\mu_1}_b U^{\mu_2}_b U^\alpha_c 
=0.
\end{aligned}
\end{equation}
Equation \eq{cont} is inconsistent with the non-degeneracy condition \eq{2ss}, and thus all $U_a^\mu$'s are parallel. 
Hence,  $U_a^\mu$ can be written as $a^\mu U_a$. 

\vspace{2mm}
{\noindent \underline{ ii). $A_{ia}^\mu = V_i^\mu U_a$}}
\vspace{2mm}

The proof in this case is parallel to that for the case $A_{ia}^\mu = V_i U_a^\mu$ shown above.
Defining $m_a$ and $n^\mu_i$ similarly as 
\begin{equation}
m_a:=\epsilon_{ij} U_a,
\qquad
n_i^\mu := \epsilon_{ij} V_j^\mu, 
\end{equation}
the condition (\ref{2sub}) becomes
\begin{equation}
n_i^\mu \left( B_{ia}-V^\beta_i \partial_\beta U_a\right)  m_a =0 \qquad (\mbox{for any}\  \mu), 
\end{equation}
where we used
\begin{equation}
n^\beta_i V^\mu_i = \epsilon_{ij} V^\beta_j V^\mu_i = -V^\beta_i n^\mu_i.
\label{nV=-Vn}
\end{equation}

As is the case $A_{ia}^\mu = V_i U_a^\mu$, 
we show $V^\mu_i = a^\mu V_i$ with proof by contradiction; 
we assume that two of $V^\mu_i$'s are not parallel to each other.
Then, two of $n_i^\mu$'s are not parallel to each other, which gives a condition that
\begin{equation}
    \bigl( B_{ia}-V^\beta_i \partial_\beta U_a\bigr)  m_a = 0~.
    \label{Bm=0}
\end{equation}
However, we can show that it is inconsistent with the non-degeneracy condition (\ref{2ss}). 
The left-hand side of (\ref{2ss}) is calculated as
\begin{eqnarray}
&&
\biggl\{
U_c^2 V_k^{\mu_1}V_k^{\mu_2} m_a n^\beta_j \left( \partial _\beta B_{jb} \right) m_b n^{\mu_3}_i  
-m_a \left[ n^{\mu_1}_j B_{jb} + n^\beta_j \partial_\beta \left( V^{\mu_1}_j U_b \right) \right] U_b V^{\mu_2}_k B_{kc} m_c n^{\mu_3}_i
\nonumber \\
&&\hspace{40mm}
+ 2 m_a \left[ n^{[\alpha|}_j B_{jb} + n^\beta_j \partial_\beta \left( V^{[\alpha|}_j U_b \right) \right] U_b V^{\mu_1}_k V^{|\mu_2]}_k U_c  \partial_\alpha \left( m_c n^{\mu_3}_i \right)
\biggr\} \xi_{\mu_1}\xi_{\mu_2}\xi_{\mu_3}
\nonumber \\
&&
=
\biggl\{
m_a V_k^{\mu_1}V_k^{\mu_2} n^\beta_j \left[ 
\left( \partial_\beta B_{jc} \right) U_b^2 m_c 
+B_{jb}U_bU_c  \partial_\beta m_c
+ \partial_\beta \left( V^{\alpha}_j U_b \right) U_b  U_c    \partial_\alpha m_c  
\right] n^{\mu_3}_i
\nonumber \\
&&\hspace{35mm}
-m_a \left[ 
n^{\mu_1}_j B_{jb} + n^\beta_j \partial_\beta \left( V^{\mu_1}_j U_b \right)
\right] U_b V^{\mu_2}_k 
\left(B_{kc} - V^\alpha_k \partial_\alpha  U_c \right) m_c n^{\mu_3}_i
\biggr\} \xi_{\mu_1}\xi_{\mu_2}\xi_{\mu_3}
\nonumber \\
&&
= 
m_a V_k^{\mu_1}V_k^{\mu_2} n^\beta_j \left[ 
 \partial_\beta\left( B_{jc} m_c \right)  U_b^2 
-B_{jb} m_b m_c \partial_\beta m_c 
+ \partial_\beta \left( V^{\alpha}_j U_b \right) U_b  U_c \partial_\alpha m_c
\right] n^{\mu_3}_i
 \xi_{\mu_1}\xi_{\mu_2}\xi_{\mu_3}
\nonumber \\
&&
=m_a V_k^{\mu_1}V_k^{\mu_2} n^\beta_j
\Bigl\{
\partial_\beta\left[V^\alpha_j \left( \partial_\alpha U_c \right) m_c \right] U_b^2 
-V^\alpha_j \left( \partial_\alpha U_b \right)  m_b m_c \partial_\beta m_c
\nonumber \\
&& \hspace{47mm}
+  \left(\partial_\beta V^{\alpha}_j \right) U_b^2  U_c \partial_\alpha m_c
+  V^{\alpha}_j \left(\partial_\beta  U_b \right)  U_b  U_c  \partial_\alpha m_c 
\Bigr\} n^{\mu_3}_i \xi_{\mu_1}\xi_{\mu_2}\xi_{\mu_3}
\nonumber \\
&&
=m_a V_k^{\mu_1}V_k^{\mu_2} n^\beta_j
\Bigl\{
  U_b^2 \left[
\left( \partial_\beta V^\alpha_j \right)  \left( \partial_\alpha U_c \right) m_c  
+V^\alpha_j  \left( \partial_\alpha U_c \right) \partial_\beta m_c
+  \left(\partial_\beta V^{\alpha}_j \right)  U_c \partial_\alpha m_c
\right]
\nonumber \\
&& \hspace{67mm}
-V^\alpha_j \left( \partial_\alpha U_b \right)\left(U_b U_c + m_b m_c \right)\partial_\beta m_c
\qquad
\Bigr\} n^{\mu_3}_i \xi_{\mu_1}\xi_{\mu_2}\xi_{\mu_3}
\nonumber \\
&&
=0,
\end{eqnarray}
where we used Eqs.~\eqref{nV=-Vn}, \eqref{Bm=0} and
\begin{equation}
\begin{aligned}
&{\cal A}^\mu_{2,ab} = m_a \left[ n^\mu_i B_{ib} + n^\beta_i \partial_\beta \left( V^\mu_i U_b \right) \right], \qquad
{\cal B}_{2,ab} = m_a n^\beta_i \partial _\beta B_{ib}, \qquad
\bar A^{2,\mu_1 \mu_2} = U_a U_a V_i^{\mu_1}V_i^{\mu_2}, 
\\
&
{\cal A}^\mu_{2,ab} m_b =0,\qquad
U_c \partial_\alpha m_c = - \left( \partial_\alpha U_c\right) m_c, \qquad 
V_j^\alpha n_j^\beta \partial_{\alpha\beta}U_c =0, \qquad 
U_a U_b+m_a m_b = U_c^2 \delta_{ab},\\
&
n^\beta_j V^{\alpha}_j \left(\partial_\beta  U_b \right)  U_b  U_c   \left( \partial_\alpha m_c  \right)
  =- n^\beta_j V^{\alpha}_j \left(\partial_\alpha  U_b \right)  U_b  U_c   \left( \partial_\beta m_c  \right).
\end{aligned}
\end{equation}
This is inconsistent with the non-degeneracy condition \eq{2ss},  and thus all $n_i^\mu$'s are parallel. 
This implies that $V_i^\mu$ can be written as 
\begin{equation}
V_j^\mu =a^\mu V_i. 
\end{equation}

\subsubsection{Further simplification of conditions (\ref{2sub}) and (\ref{2ss})}
\label{sec:2simfin}

We have shown that, for the invertibility conditions to be satisfied, $A_{ia}^\mu$ should be written as 
\begin{equation}
A_{ia}^\mu = a^\mu V_i U_a . 
\end{equation}
Without loss of generality, we can normalize $V_i$ and  $U_a$ as $V_i V_i=1=U_aU_a$. 
We define unit vectors $n_i$ and $m_a$ that are normal to $V_i$ and $U_a$ respectively as 
\begin{equation}
n_i:=\epsilon_{ij} V_j,
\qquad
m_a := \epsilon_{ab} U_b. 
\end{equation}
Since $A_{ia}^\mu$ is written with $a^\mu$, $V_i$ and  $U_a$, the matrices $\bar A_{ai}^\mu$, $\tilde A_{ai}^\mu$, also ${\bar A}^{2,\mu\nu}$, ${\cal A}_{2,ab}^\mu$ and ${\cal B}_{2,ab}$ are written as
\begin{eqnarray}
&&\bar A_{ai}^\mu = a^\mu m_a n_i, \qquad
\tilde A_{ai}^\mu= a^\mu U_a V_i, \qquad
{\bar A}^{2,\mu\nu}=a^\mu a^\nu, \nonumber\\
&&
{\cal A}_{2,ab}^\mu = a^\mu m_a n_i \left[ B_{ib}+ a^\beta \left( \partial_\beta V_i\right) U_b \right],
\qquad
{\cal B}_{2,ab} = m_a n_i a^\beta \partial_\beta B_{ib}.
\end{eqnarray}
Substituting the above equations into the condition (\ref{2sub}) 
we have 
\begin{equation}
n_i B_{ia} m_a =0. \label{nBm}
\end{equation}
The non-degeneracy condition (\ref{2ss}) becomes
\begin{equation}
n_i a^\beta\left( \partial_\beta B_{ia} \right) m_a - n_i B_{ia}U_a V_j B_{jb}m_b 
- n_i a^\beta\left( \partial_\beta V_i \right) V_j B_{ja}m_a \neq 0.
\label{nond}
\end{equation}
The first term of the left-hand side can be transformed as
\begin{align}
n_i a^\beta\left( \partial_\beta B_{ia} \right) m_a 
&= - a^\beta\left( \partial_\beta n_i \right) B_{ia} m_a -n_i B_{ia} a^\beta \partial_\beta m_a  \nonumber\\
&= - a^\beta\left( \partial_\beta n_i \right) B_{ia} m_a -n_i B_{ia} U_a U_b a^\beta \partial_\beta m_b \nonumber\\
&= - a^\beta\left( \partial_\beta n_i \right) B_{ia} m_a +n_i B_{ia} U_a a^\beta\left( \partial_\beta U_b \right)  m_b,
\end{align}
where we use the fact that $U_a U_b +m_a m_b = \delta_{ab}$ and Eq.~(\ref{nBm}). 
The last term of \eq{nond} can be transformed as
\begin{equation}
- n_i a^\beta\left( \partial_\beta V_i \right) V_j B_{ja}m_a = a^\beta\left( \partial_\beta n_i \right) B_{ia} m_a,
\end{equation}
where we use the fact that $V_i V_j +n_i n_j = \delta_{ij}$ and Eq.~(\ref{nBm}).
Then, condition \eq{nond} can be written as
\begin{equation}
 n_i B_{ia}U_a  \left( V_j B_{jb} -  a^\beta \partial_\beta U_b \right) m_b \neq 0.
\end{equation}
This means that both of
\begin{equation}
 n_i B_{ia}U_a\neq 0~, \qquad
 \left( V_j B_{jb} -  a^\beta \partial_\beta U_b \right) m_b \neq 0 \label{nBU}
\end{equation}
should be satisfied.
As a result, the invertibility conditions (\ref{detA}), (\ref{2sub}) and (\ref{2ss}) are equivalent to the simplified conditions 
(\ref{aVU}), (\ref{nBm}) and (\ref{nBU}).

\section{Sufficiency of the invertibility conditions}\label{Sec:Suf}

In the previous section, we derived the necessary conditions 
(\ref{aVU}), (\ref{nBm}) and (\ref{nBU})
for invertibility of a field transformation involving two fields and up to first-order derivatives of them. 
In this section, we show that these conditions are also sufficient conditions, that is, 
(\ref{aVU}), (\ref{nBm}) and (\ref{nBU}) are the necessary and sufficient conditions for the invertibility in the two-field case. 
As a preliminary step, in section \ref{PI}, we introduce a notion of ``partial invertibility'' for a field transformation
whose inverse transformation is uniquely determined for a part of variables. 
Then, we show in section \ref{subSC} with the partial invertibility that (\ref{aVU}), (\ref{nBm}) and (\ref{nBU}) are the necessary and sufficient conditions for invertibility.\footnote{See Appendix~\ref{App:implicit-function} for the difference between standard approach based on the inverse function theorem on functional spaces and our approach based on the implicit function theorem on finite-dimensional subspaces associated with the functional space.}

\subsection{Partial invertibility} \label{PI}

We consider invertibility of a transformation 
$\psi_a \to \phi_i$.
Suppose that 
the transformation can be described by
\begin{equation}
F_I(\psi_a, \partial_\mu \psi_a, \partial_\mu \partial_\nu \psi_a, \cdots;  \phi_i, \partial_\mu \phi_i, \partial_\mu \partial_\nu \phi_i, \cdots; x^\mu) =0,
\label{F_I=0}
\end{equation}
where $I$ runs from $1$ to a constant ${\cal N}$. 
If ${\cal N}$ is equal to the number of all the degrees of freedom in $\psi_a, \partial_\mu \psi_a, \partial_\mu \partial_\nu \psi_a, \cdots$,%
\footnote{
For instance, if $F_I$ depends on up to the second order derivative of $\psi_a$, the number of all the degrees of freedom of 
$(\psi_a, \partial_\mu \psi_a, \partial_\mu \partial_\nu \psi_a)$ 
is 
$N+N\times D +N \times D(D+1)/2$,
where $N$ and $D$ are the number of $\psi_a$ fields and the dimension of spacetime, respectively.} 
which is denoted by ${\cal N}_\psi$,
we can use the implicit function theorem by regarding $\psi_a, \partial_\mu \psi_a, \partial_\mu \partial_\nu \psi_a, \cdots$ as independent variables at a point in spacetime.
However, we may not need to
have ${\cal N_\psi}$ equations of the form of \eqref{F_I=0} to fix $\psi_a$ uniquely in terms of $\phi_a$,
indeed,
even if we have less than ${\cal N_\psi}$ equations, {\it i.e.} ${\cal N}<{\cal N_\psi}$; it may be possible to prove the uniqueness of $\psi_a$ if the equations have a special structure.
Once $\psi_a$ is fixed in this manner, its derivatives $\partial_\mu \psi_a, \partial_\mu\partial_\nu \psi_a, \ldots $ are uniquely fixed by
differentiating $\psi_a$ repeatedly.

To illustrate the procedure explained above, 
we first discuss a generic case, where equations are for functions of variables $x_a$, $y_b$, $z_c$ and $w_c$,
\begin{equation}
F_I(x_a, y_b, z_c; w_d) =0,
\label{xyzw}
\end{equation}
where $x_a$ corresponds to $\psi_a$, while $y_b$ and $z_c$ correspond to some combinations of the derivatives $\partial_\mu \psi_a, \partial_\mu \partial_\nu\psi_a, \cdots$.\footnote{
For our purpose to prove the invertibility for the two-field case with first-order derivative
given by Eq.~(\ref{redef}), 
$x_a$ corresponds to $\psi_a$,  $y_b$ is a component of $\partial_\mu \psi_a$, and $z_c$ correspond to the other components of the derivative $\partial_\mu \psi_a$ and $\partial_\mu \partial_\nu \psi_a$.}
The difference between $y_b$ and $z_c$ is explained later.
$w_d$ corresponds to the other variables $\phi_i$ and their derivatives  $\partial_\mu \phi_i, \partial_\mu \partial_\nu \phi_i, \cdots$.
If the number of equations is the same as sum of those of $x_a$, $y_b$ and $z_c$, we can use the implicit function theorem for them, {\it i.e.} $x_a$, $y_b$ and $z_c$ can be fixed uniquely
in terms of $w_d$,
provided that the conditions for the implicit function theorem are satisfied.
However, even if the number of equations is less than the sum of those of $x_a$, $y_b$ and $z_c$,
it may be possible to fix $x_a$ in terms $w_d$ uniquely
if Eq.~\eqref{xyzw} have a certain structure, as explained below.
Some of $y_b$ and $z_c$ may not be fixed uniquely in this process, but the uniqueness of $x_a$ can be shown independently.

The variables $x_a$ are fixed by stepwise application of the inverse function theorem provided that the function $F_I$ has a structure given by Eq.~\eqref{yzcomb}, which will be introduced shortly.
Suppose that we have $N+M$ equations of the form of Eq.~\eqref{xyzw}, {\it i.e.} $I$ runs from $1$ to $N+M(={\cal N})$, where $N$ denotes the number of the fields $x_a$ 
and $M$ is a positive integer. 
The variation of \eq{xyzw} is 
\begin{align}
\delta F_I(x_a, y_b, z_c; w_d)
&=
\frac{\partial F_I}{\partial x_a}(x_a, y_b, z_c; w_d) \delta x_a 
+\frac{\partial F_I}{\partial y_b}(x_a, y_b, z_c; w_d) \delta y_b
\nonumber \\ & ~
+\frac{\partial F_I}{\partial z_c}(x_a, y_b, z_c; w_d) \delta z_c
+\frac{\partial F_I}{\partial w_d}(x_a, y_b, z_c; w_d) \delta w_d
~.
\end{align}
Here, we assume that $\delta y_b$ and $\delta z_c$ appear only as $M$ independent combinations in any $\delta F_I$, 
{\it i.e.} $\delta F_I$ is expressed as 
\begin{align}
\delta F_I(x_a, y_b, z_c; w_d)
&=  \frac{\partial F_I}{\partial x_a}(x_a, y_b, z_c; w_d) \delta x_a 
+\frac{\partial F_I}{\partial w_d}(x_a, y_b, z_c; w_d) \delta w_d
\nonumber \\ & ~
+\sum_{b=1}^M B_{Ib} (x_a, y_b, z_c; w_d) \left( \delta y_b + \tilde B_{bc}(x_a, y_b, z_c; w_d) \delta z_c \right)~.
\label{yzcomb}
\end{align}
Without loss of generality, the principal components of $M$ independent combinations are set to be $y_b$, 
which means that the rank of $B_{Ib}$ is $M$. 
This makes the difference between $y_b$ and $z_c$.
It also fixes the number of $y_b$ to be $M$.  
Then, picking up $M$ equations from $F_I$'s such that $y_b$ is fixed uniquely, we use the implicit function theorem to fix only $y_b$. 
Without loss of generality,
we assume that these equations are given by $F_I$ with $N+1\leq I \leq N+M$,
that is, the square matrix $B_{Ib}$ with $I = N+1,\dots,  N+M$ and $b=1,\dots, M$ is regular.
Since the number of these equations and that of $y_b$ are the same, we can apply the implicit function theorem for $y_b$, 
and then, $y_b$ is uniquely expressed in terms of $x_a$, $z_c$ and $w_d$.

Now, $y_b$'s are uniquely written as functions of $x_a$, $z_c$ and $w_d$. 
Substituting $y_b(x_a,z_c,w_d)$ into Eq.~\eq{xyzw}  for $N+1\le I \le N+M$,
the derivative of Eq.~\eq{xyzw} with respect to $z_c$ is given by
\begin{equation}
\sum_{b=1}^M B_{Ib} (x_a, y_b, z_c; w_d) \left( \frac{\partial y_b}{\partial z_c} + \tilde B_{bc} \right) = 0.
\label{sumB}
\end{equation}
Since $B_{Ib}$ is assumed to be regular, this equation implies
\begin{equation}
\frac{\partial y_b}{\partial z_c} + \tilde B_{bc} =0 ~.
\label{y}
\end{equation}
This equation fixes the $z_c$ dependence of $y_b(x_a,z_c,w_d)$.
Now the remaining equations can be written purely in terms of $x_a$, $z_c$ and $w_d$ as
\begin{equation}
F_I\bigl(x_a, y_b (x_a, z_c, w_d), z_c; w_d\bigr) =0~,
\end{equation}
where $I$ runs from 1 to $N$. 
Their variation is given by
\begin{align}
\delta F_I\bigl(x_a, y_b (x_a, z_c, w_d), z_c; w_d\bigr)
&
=
\Biggl[\frac{\partial F_I}{\partial x_a}(x_a, y_b, z_c; w_d) \delta x_a 
+\frac{\partial F_I}{\partial w_d}(x_a, y_b, z_c; w_d) \delta w_d
\nonumber \\ & \quad
+\sum_{b=1}^M B_{Ib} (x_a, y_b, z_c; w_d) 
\left( \frac{\partial y_b}{\partial x_a} \delta x_a  
+\frac{\partial y_b}{\partial w_d} \delta w_d
\right)
\Biggr]_{y_b=y_b (x_a, z_c, w_d)}.
\end{align}
The $\delta z_c$ term is canceled because of Eq.~\eq{y} in this equation.
This implies that $F_I$'s for $I=1,\dots,N$ are independent of $z_c$, {\it i.e.} they are functions of only $x_a$ and $w_d$. 
Since the number of equations is the same as that of $x_a$, we can use the implicit function theorem;
if 
\begin{equation}
\det \left[\frac{\partial F_I}{\partial x_a}(x_a, y_b, z_c; w_d) 
+\sum_{b=1}^M B_{Ib} (x_a, y_b, z_c; w_d) 
 \frac{\partial y_b}{\partial x_a}  
\right]_{y_b=y_b (x_a, z_c, w_d)}
\neq 0
\end{equation}
is satisfied,
$x_a$ has a unique solution locally and it is written in terms of only $w_d$. 

In the next section, we will see that the transformation \eq{redef} in the two field case behaves as Eq.~\eqref{yzcomb} once the invertibility conditions (\ref{aVU}), (\ref{nBm}) and (\ref{nBU}) are imposed, and then $\psi_i$ is fixed in terms of $\phi_a$ uniquely as a consequence.

\subsection{Sufficient conditions for invertibility of our transformation} \label{subSC}

The transformation equation \eq{redef} can be rewritten in terms of a function of $\psi_a$, $\partial_\mu \psi_a$ and  $\phi_i$ as
\begin{equation}
F_i(\psi_a,\partial_\mu \psi_a; \phi_i ; x^\mu):= \bar \phi_i (\psi_a,\partial_\mu \psi_a;  x^\mu) - \phi_i =0.
\label{F_Idef}
\end{equation}
We analyze this equation in the two field case imposing the necessary conditions for the invertibility, Eqs.~(\ref{aVU}), (\ref{nBm}) and (\ref{nBU}).
Operating $n_i \partial_\mu$ on Eq.~\eqref{F_Idef}, we have 
\begin{equation}
G_\mu (\psi_a,\partial_\mu \psi_a; \phi_i, \partial_\mu \phi_i ; x^\mu) := n_i \partial_\mu F_i =
n_i B_{ib} U_b U_a \partial_\mu \psi_a - n_i C_{i\mu } - n_i\partial_\mu \phi_i =0,
\end{equation}
where 
$B_{ib}$, $n_i$ and $U_a$ are defined in section \ref{sec:2sim} and $C_{i \mu}:= \partial \bar \phi_i/ \partial x^\mu$.
The variation of $F_i$ becomes
\begin{align}
n_i \delta F_i&=(n_i B_{ib} U_b) U_a \delta \psi_a ~ , \label{nF}\\
V_i \delta F_i&=(V_i B_{ib} U_b) U_a \delta \psi_a + (V_i B_{ib} m_b) m_a \delta \psi_a + a^\mu U_a \delta (\partial_\mu \psi_a) ~,
\label{VF}
\end{align}
where we omit the variation with respect to $\phi_i$,
because it is irrelevant to the condition for the implicit function theorem. 

Now let us evaluate the variation of $a^\mu G_\mu$, which is given as 
\begin{align}
\delta \left(a^\mu G_\mu \right)
&=
a^\mu n_i \delta \left(\partial_\mu F_i \right)  = n_i \delta \left(\partial_\mu \bar \phi_i \right)
\nonumber \\ 
&=
a^\mu n_i \delta\left(
\frac{\partial \bar \phi_i}{\partial \left(\p_\alpha \psi_a\right) } \p_\mu \p_\alpha \psi_a
+\frac{\partial \bar \phi_i}{\partial  \psi_a } \p_\mu  \psi_a
+\frac{\partial \bar \phi_i}{\partial x^\mu } \right)
\nonumber \\ 
&=
a^\mu n_i
\biggl[
\left(
\frac{\partial^2 \bar \phi_i}{\partial \left(\p_\alpha \psi_a\right) \partial \left(\p_\beta \psi_b\right) } \p_\mu \p_\alpha \psi_a 
+\frac{\partial^2 \bar \phi_i}{\partial  \psi_a \partial \left(\p_\beta \psi_b \right)} \p_\mu  \psi_a
+\frac{\partial^2 \bar \phi_i}{\partial x^\mu  \partial\left( \p_\beta \psi_b\right)} \right)
\delta \left(\p_\beta \psi_b
\right)
\nonumber \\ 
& \hspace{17mm} 
+\left( \frac{\partial^2 \bar \phi_i}{\partial \left(\p_\alpha \psi_a\right)\partial  \psi_b } \p_\mu \p_\alpha \psi_a
+\frac{\partial^2 \bar \phi_i}{\partial  \psi_a \partial \psi_b } \p_\mu  \psi_a
+\frac{\partial^2 \bar \phi_i}{\partial x^\mu  \partial \psi_b} \right)\delta  \psi_b 
+B_{ia } \delta \left(\p_\mu  \psi_a \right)
\biggr]
\nonumber \\ 
&= a^\mu  n_i (\partial_\mu B_{ib})U_b U_a \delta \psi_a
+a^\mu n_i (\partial_\mu B_{ib})m_b m_a \delta \psi_a + n_i ( a^\mu \partial_\mu V_i + B_{ib}U_b) a^\nu U_a \delta (\partial_\nu \psi_a), \label{aG}
\end{align}
where we used $\partial_\mu F_i=0$ at the first equality, and some formulae in section~\ref{sec:2simfin} such as the degeneracy conditions $n_i  \frac{\partial \bar \phi_i}{\partial (\p_\alpha \psi_a) } =  n_i A^\alpha_{ia}=0$, $n_iB_{ia}m_a=0$ at the fourth equality.
The key point is that,
in Eqs.~\eq{nF}, \eq{VF} and \eq{aG}, $\delta (\partial_\mu \psi_a)$ appears only in a linear combination $a^\mu U_a \delta (\partial_\mu \psi_a)$. 
The last term proportional to $a^\mu U_a \delta (\partial_\mu \psi_a)$ in Eq.~(\ref{aG}) 
corresponds to the last line of Eq.~(\ref{yzcomb}) for $M=1$. 
This allows us to use
the result of section~\ref{PI} by regarding $\psi_a$, $\partial_\mu \psi_a$ and $\partial_\mu \partial_\nu \psi_a$ as independent variables and also identifying $x_a$, $y_b$ and $z_c$ as $\psi_a$, a component of $\partial_\mu \psi_a$ and the other components of ($\partial_\mu \psi_a$, $\partial_\mu \partial_\nu \psi_a$), respectively. 
Then, the condition of invertibility is
\begin{align}
0 &\neq
\det
\left(
\begin{array}{cc:c}
  V_i \dfrac{\partial F_i}{\partial \psi_a} U_a &
  V_i \dfrac{\partial F_i}{\partial \psi_a} m_a &
  V_i \dfrac{\partial F_i}{\partial \left(\partial_\nu\psi_a\right)} U_a
\\
  n_i \dfrac{\partial F_i}{\partial \psi_a} U_a &
  n_i \dfrac{\partial F_i}{\partial \psi_a} m_a &
  n_i \dfrac{\partial F_i}{\partial \left(\partial_\nu\psi_a\right)} U_a
\\ \hdashline
 \dfrac{\partial \left(a^\mu G_\mu \right)}{\partial \psi_a}U_a &
 \dfrac{\partial \left(a^\mu G_\mu \right)}{\partial \psi_a}m_a &
 \dfrac{\partial \left(a^\mu G_\mu \right)}{\partial \left(\partial_\nu \psi_a\right)} U_a\\
\end{array}
\right)
\nonumber \\ 
&=
\det
\left(
\begin{array}{cc:c}
V_i B_{ia} U_a & V_iB_{ia}m_a & a^\nu  \\
n_i B_{ia} U_a & 0 & 0  \\ \hdashline
a^\mu n_i\left(\partial_\mu B_{ia}\right)U_a & 
a^\mu n_i\left(\partial_\mu B_{ia}\right)m_a &
n_i\left(a^\mu \partial_\mu V_i + B_{ia}U_a \right) a^\nu
\end{array}
\right)
\nonumber \\
&=
-a^\nu (n_iB_{ia}U_a)^2(V_j B_{jb}-a^\mu\partial_\mu U_b) m_b~,
\label{nBU_sufficient}
\end{align}
where the components of the matrix are expressed with respect to the bases $(V_i, n_i)$, $(U_a, m_a)$,
and also 
the degenerate row and column whose components are completely zero are removed.
Equation~\eqref{nBU_sufficient} is satisfied if $n_iB_{ia}U_a \neq 0$ and $(V_j B_{jb}-a^\mu\partial_\mu U_b) m_b\neq 0$.
Therefore, the necessary conditions for the invertibility, Eqs.~(\ref{aVU}), (\ref{nBm}) and (\ref{nBU}), are also the sufficient conditions.

\section{Examples of invertible transformations}
\label{sec:examples}

Invertibility conditions for transformations $\phi_i = \phi_i(\psi_a, \partial \psi_a)$ are established in the previous sections. 
In this section, we propose some non-trivial examples of field transformations that satisfy the invertibility conditions.
In principle, the most general invertible transformation could be constructed by finding the general solution of the invertibility conditions.
Unfortunately it is not an easy task,\footnote{See Ref.~\cite{Babichev:2019twf} for a previous attempt of constructing examples of invertible transformations.} so we proceed with introducing the following ansatz of field transformations:
\begin{enumerate}
\item $\displaystyle \phi_i = b(\psi_a, Y_\mu) V_i(\psi_a) + \tilde V_i(\psi_a)
\qquad
\left(Y_\mu \equiv U_a(\psi_a)\partial_\mu \psi_a \right)$
\item $\displaystyle \phi_i = b (\psi_a, Y_a ) V_i(\psi_a) + \tilde V_i(\psi_a)
\qquad
\left( Y_a \equiv \tilde a^\mu(\psi_a) \partial_\mu \psi_a \right)$
\item $\displaystyle \phi_i = \phi_i ( \psi_a , Y )
\hspace{33.0mm}
\left( Y \equiv \tilde a^\mu (\psi_a) U_a(\psi_a) \partial_\mu \psi_a \right)$
\end{enumerate}
These ansatze are chosen so that the first degeneracy condition (\ref{aVU}) among the invertibility conditions is automatically satisfied.
The second degeneracy condition (\ref{nBm}) and the non-degeneracy conditions (\ref{nBU}) are then used to constrain the form of the transformations.
Although these transformations may not span the most general transformation satisfying the invertibility conditions, they would give a good starting point for investigating the most-general invertible field transformation.

In the derivation below, we use the fact that the vector $U_a$ can be set to a constant vector $(0,1)$ 
if it depends only on $\psi_a$.
We explain this method in Appendix~\ref{App:Setting-U}.

\subsection{Ansatz 1: $\displaystyle \phi_i = b(\psi_a,  U_a(\psi_a)\partial_\mu \psi_a) V_i(\psi_a) + \tilde V_i(\psi_a)$}

We first consider the following ansatz for a field transformation given by
\begin{equation}
\phi_i = b(\psi_a, Y_\mu) V_i(\psi_a) + \tilde V_i(\psi_a)
\qquad
\left(Y_\mu \equiv U_a(\psi_a)\partial_\mu \psi_a \right)~,
\label{ex1}
\end{equation}
where $V_i$ is normalized as $V_i V_i = 1$.
In this transformation, $U_a$ can be set to $U_a = (0,1)$ without loss of generality.
It then follows 
\begin{equation}
U_a = (0,1),
\quad
m_a = (1,0),
\quad
n_i = \epsilon_{ij} V_j~.
\label{ex1_setU}
\end{equation}

For (\ref{ex1}), the first degeneracy condition~(\ref{aVU}) is given by
\begin{equation}
A^\mu_{ia}
= \frac{\partial \phi_i}{\partial\left(\partial_\mu \psi_a\right)}
= \frac{\partial b}{\partial Y_\mu} U_a V_i
\equiv a^\mu U_a V_i~.
\end{equation}
The second degeneracy condition~(\ref{nBm}) is evaluated as follows.
\begin{gather}
B_{ia}
= \frac{\partial \phi_i}{\partial \psi_a}
=
\frac{\partial b}{\partial \psi_a} V_i 
+ b \frac{\partial V_i}{\partial \psi_a} 
+ \frac{\partial \tilde V_i}{\partial \psi_a} ~,
\\
n_i B_{ia} m_a
=
b(\psi_a, Y_\mu) n_i \frac{\partial V_i}{\partial \psi_1} 
+ n_i\frac{\partial \tilde V_i}{\partial \psi_1}~.
\label{ex1_deg2}
\end{gather}
In (\ref{ex1_deg2}), only $b$ depends on $Y_\mu$ while the other terms depend only on $\psi_a$.
Because Eq.~(\ref{ex1_deg2}) should be satisfied identically for any $\psi_a$ and $Y_\mu$, it implies 
\begin{align}
b\,n_i \frac{\partial V_i}{\partial \psi_1} 
&=0~,
\label{ex1_deg2_1}
\\
n_i\frac{\partial \tilde V_i}{\partial \psi_1} &= 0~.
\label{ex1_deg2_2}
\end{align}
Equation~(\ref{ex1_deg2_1})
implies either $n_i \frac{\partial V_i}{\partial \psi_1} = 0$ or $b=0$. $n_i \frac{\partial V_i}{\partial \psi_1} = 0$ implies $V_i = V_i(\psi_2)$ thanks to the normalization $V_i V_i=1$, 
while $b= 0$ gives a transformation without derivatives $\phi_i = \phi_i(\psi_a)$. 
Below we focus on the case $n_i \frac{\partial V_i}{\partial \psi_1} = 0$, in which the transformation depends on derivatives $\phi_i = \phi_i(\psi_a,\partial\psi_a)$.
In this case, Eq.~(\ref{ex1_deg2_2}) implies that $\partial \tilde V_i / \partial \psi_1$ is parallel to $V_i$, that is,
\begin{equation}
\frac{\partial \tilde V_i}{\partial \psi_1} = \tilde c(\psi_a) V_i(\psi_2)
\quad \Rightarrow\quad
\tilde V_i  = c(\psi_a)V_i(\psi_2) + \hat V_i(\psi_2)
\label{ex1_tildeV}
\end{equation}
for some functions $\tilde c(\psi_a)$, $c(\psi_1) = \int \tilde c(\psi_a) d\psi_1$ and $\hat V_i(\psi_2)$. The $c\,V_i$ term can be absorbed into the $b\, V_i$ term in the ansatz (\ref{ex1}), and the remainder $\hat V_i(\psi_2)$ depends only on $\psi_2$. 
It implies that, 
once Eq.~(\ref{ex1_setU}) is imposed,
we may set $\tilde V_i = \tilde V_i(\psi_2)$ in the ansatz~(\ref{ex1}) without loss of generality by absorbing the $c\, V_i$ term.
Below we continue the derivation under this assumption.

The non-degeneracy conditions~(\ref{nBU})
are given by
\begin{align}
0 &\neq 
n_i B_{ia}U_a 
=
b \, n_i\frac{\partial V_i}{\partial \psi_2} + n_i\frac{\partial \tilde V_i}{\partial \psi_2}~,
\label{ex1_nondeg-1}
\\
0 &\neq
\left(V_i B_{ia} - a^\mu \partial_\mu U_a \right)m_a
= 
\left(
\frac{\partial b}{\partial \psi_a} 
+ \frac{\partial b}{\partial Y_\mu} \partial_\mu \psi_b \frac{\partial U_b}{\partial\psi_a}
+ V_i \frac{\partial \tilde V_i}{\partial \psi_a}
- a^\mu \partial_\mu U_a
\right)m_a
=
\frac{\partial b}{\partial \psi_1} 
~,
\label{ex1_nondeg-2}
\end{align}
which can be regarded as constraints on $b$, $V_i$ and $\tilde V_i$.

To summarize, a transformation given by
\begin{equation}
\phi_i = b(\psi_a, Y_\mu) V_i(\psi_2) 
+ \tilde V_i(\psi_2)
\qquad
\left(Y_\mu \equiv \partial_\mu \psi_2 \right)~.
\label{ex1_final}
\end{equation}
is invertible if 
Eqs.~(\ref{ex1_nondeg-1}) and (\ref{ex1_nondeg-2})
are satisfied. One may further apply an invertible transformation $\psi_a = \psi_a (\tilde \psi_b)$ 
such that $\det(\partial\psi_a/\partial\tilde\psi_b)\neq 0$
on Eq.~(\ref{ex1_final}) to re-introduce nontrivial $U_a(\tilde \psi_a)$ and construct a transformation of the form of Eq.~(\ref{ex1}).

\subsection{Ansatz 2: $\displaystyle \phi_i = b\left(\psi_a, \tilde a^\mu(\psi_a) \partial_\mu \psi_a\right) V_i(\psi_a) + \tilde V_i(\psi_a)$}

The next ansatz is given by 
\begin{equation}
\phi_i = b\left(\psi_a, Y_a\right) V_i(\psi_a) + \tilde V_i(\psi_a)
\qquad
\left(
Y_a \equiv \tilde a^\mu(\psi_a) \partial_\mu \psi_a
\right)~.
\label{ex2}
\end{equation}
The first degeneracy condition~(\ref{aVU}) is automatically satisfied for this ansatz:
\begin{equation}
A_{ia}^\mu = \tilde a^\mu \frac{\partial b}{\partial Y_a} V_i
\equiv a^\mu(\psi_a, Y_a) U_a(\psi_a, Y_a) V_i(\psi_a)~.
\end{equation}
The second degeneracy~(\ref{nBm}) condition is given as
\begin{gather}
B_{ia} =
\frac{\partial b}{\partial \psi_a} V_i 
+ b\frac{\partial V_i}{\partial \psi_a} 
+ \frac{\partial \tilde V_i}{\partial \psi_a}~,
\\
\therefore \quad
n_i B_{ia} m_a
=
n_i 
\left(
 b\frac{\partial V_i}{\partial \psi_a} 
+ \frac{\partial \tilde V_i}{\partial \psi_a}
\right)m_a
\propto 
n_i 
\left(
 b\frac{\partial V_i}{\partial \psi_a} 
+ \frac{\partial \tilde V_i}{\partial \psi_a}
\right)
\epsilon_{ab}
\frac{\partial b}{\partial Y_b}
= 0.
\label{ex2_deg2}
\end{gather}

Equation (\ref{ex2_deg2}) is equivalent to
\begin{gather}
\frac{n_i 
\left(
 b\frac{\partial V_i}{\partial \psi_1} 
+ \frac{\partial \tilde V_i}{\partial \psi_1}
\right)}{
n_i 
\left(
 b\frac{\partial V_i}{\partial \psi_2} 
+ \frac{\partial \tilde V_i}{\partial \psi_2}
\right)
}
=
\frac{\frac{\partial b}{\partial Y_1}}{\frac{\partial b}{\partial Y_2}}
\equiv -c\left(
\psi_a, Y_a
\right)
\label{ex2_deg2-0}
\\
\Leftrightarrow\quad
n_i
\left(
 b\frac{\partial V_i}{\partial \psi_1} 
+ \frac{\partial \tilde V_i}{\partial \psi_1}
\right)
+ c\left(
\psi_a, Y_a
\right)
n_i
\left(
 b\frac{\partial V_i}{\partial \psi_2} 
+ \frac{\partial \tilde V_i}{\partial \psi_2}
\right) = 0~,
\quad
\frac{\partial b}{\partial Y_1} + c(\psi_a,Y_a)
\frac{\partial b}{\partial Y_2}=0.
\label{ex2_deg2-1}
\end{gather}
Equation~(\ref{ex2_deg2-1}) may be solved for $\tilde V_i(\psi_a)$ and $b(\psi_a, Y_a)$ once $V_i(\psi_a)$ and $c(\psi_a,Y_a)$ are freely specified, and each solution gives an invertible transformation as long as it satisfies the non-degeneracy conditions~(\ref{nBU}).

In a special case where the function $c(\psi_a, Y_a)$ depends only on $\psi_a$ but not on $Y_a$,
one can construct invertible transformations more explicitly as follows.
When $c=c(\psi_a)$,
we may apply an invertible transformation $\psi_a = \psi_a(\tilde \psi_a)$ on Eq.~(\ref{ex2_deg2-1}) to eliminate $c(\psi_a)$, that is, Eq.~(\ref{ex2_deg2-1}) transforms under $\psi_a = \psi_a(\tilde \psi_a)$ as
\begin{align}
0&=
\frac{\partial \tilde \psi_a}{\partial \psi_1}
n_i
\left(
 b\frac{\partial V_i}{\partial \tilde \psi_a} 
+ \frac{\partial \tilde V_i}{\partial \tilde \psi_a}
\right)
+ c(\psi_a)
n_i
\frac{\partial \tilde \psi_a}{\partial \psi_2}
\left(
\psi_a, Y_a
\right)
\left(
 b\frac{\partial V_i}{\partial \tilde \psi_a} 
+ \frac{\partial \tilde V_i}{\partial \tilde \psi_a}
\right) 
\notag \\
&=
\left(
\frac{\partial \tilde \psi_1}{\partial \psi_1}
+ c(\psi_a) \frac{\partial \tilde \psi_1}{\partial \psi_2}
\right)
n_i
\left(
 b\frac{\partial V_i}{\partial \tilde \psi_1} 
+ \frac{\partial \tilde V_i}{\partial \tilde \psi_1}
\right) 
+\left(
\frac{\partial \tilde \psi_2}{\partial \psi_1}
+ c(\psi_a) \frac{\partial \tilde \psi_2}{\partial \psi_2}
\right)
n_i
\left(
 b\frac{\partial V_i}{\partial \tilde \psi_2} 
+ \frac{\partial \tilde V_i}{\partial \tilde \psi_2}
\right)~,
\label{ex2_deg2-1mod}
\end{align}
and then the function $\tilde \psi_a = \tilde \psi_a(\psi_a)$ may be chosen (at least locally) so that
\begin{equation}
\frac{\partial \tilde \psi_2}{\partial \psi_1}
+ c(\psi_a) \frac{\partial \tilde \psi_2}{\partial \psi_2}
= 0~,
\end{equation}
with which Eq.~(\ref{ex2_deg2-1mod}) reduces to 
\begin{equation}
n_i
\left(
 b\frac{\partial V_i}{\partial \tilde \psi_1} 
+ \frac{\partial \tilde V_i}{\partial \tilde \psi_1}
\right)
 = 0~.
\end{equation}
This is equivalent to setting $c(\psi_a)= 0$ in Eq.~(\ref{ex2_deg2-1}) by means of a transformation in $\psi_a$ space.

When $c(\psi_a)=0$, Eq.~(\ref{ex2_deg2-0}) implies that 
\begin{equation}
n_i(\psi_a)\left(
b(\psi_a, Y_a)
\frac{\partial V_i(\psi_a)}{\partial \psi_1} 
+ \frac{\partial \tilde V_i(\psi_a)}{\partial\psi_1} 
\right)
= 0,
\qquad
\frac{\partial b}{\partial Y_1} = 0~.
\end{equation}
The second equation implies $b = b(\psi_a, Y_2)$,
which gives $U_a = (0,1)$ and $m_a = (1,0)$.
In the first equation, the first term involving $b(\psi_a, Y_2)$ depends on $Y_2$ while the second term is independent of $Y_a$, then it follows that 
\begin{align}
b(\psi_a, Y_2)
n_i(\psi_a)
\frac{\partial V_i(\psi_a)}{\partial  \psi_1} &=0 ,
\label{ex2_deg2-2-1}
\\
n_i(\psi_a)
\frac{\partial \tilde V_i(\psi_a)}{\partial  \psi_1}&=0 ~ .
\label{ex2_deg2-2-2}
\end{align}
Equation~(\ref{ex2_deg2-2-1}) implies either $V_i= V_i(\psi_2)$ or $b=0$, for the latter of which the transformation (\ref{ex2}) does not involve $\partial \psi_a$. 
Below we focus on the former case, which gives a transformation with derivatives.
In this case,
we can show that 
$\tilde V_i  = f(\psi_a) V_i(\psi_2) + \hat V_i(\psi_2)$ for some function $f(\psi_a)$ and $\hat V_i(\psi_2)$ using Eq.~(\ref{ex2_deg2-2-2}) as we did around Eq.~(\ref{ex1_tildeV}) for the previous ansatz. Then, once setting $U_a=(0,1)$, by absorbing the $c\, V_i$ term into the $b\,V_i$ term we may set $\tilde V_i = \tilde V_i(\psi_2)$ without loss of generality.

To summarize the results above,
a transformation
\begin{equation}
\phi_i = b(\psi_a, Y_2) V_i(\psi_2) + \tilde V_i(\psi_2)
\qquad
\left(
Y_a \equiv \tilde a^\mu(\psi_a) \partial_\mu \psi_a
\right)
\label{ansatz2final}
\end{equation}
may be invertible if Eq.~(\ref{ex2_deg2-2-2}) is satisfied.
Adding to that, the  non-degeneracy conditions~(\ref{nBU}) must be satisfied for the invertibility, and also 
an invertible transformation $\psi_a = \psi_a (\tilde \psi_b)$ 
may be applied to construct a transformation with nontrivial $U_a(\psi_a)$.
Actually this transformation is a special case of the transformation (\ref{ex1_final}) examined in the previous section, although we have started from a different ansatz.
This result follows from the assumption $c=c(\psi_a)$ imposed at Eq.~(\ref{ex2_deg2-1mod}), and more general transformations are obtained if we solve Eq.~(\ref{ex2_deg2-1}) for $c=c(\psi_a,Y_a)$.

\subsection{Ansatz 3: $\displaystyle \phi_i = \phi_i\left(
\psi_a , \tilde a^\mu (\psi_a) U_a(\psi_a) \partial_\mu \psi_a
\right)$}

The third ansatz we consider is 
\begin{equation}
\phi_i = \phi_i\left(
\psi_a , Y
\right)
\qquad
\left(
Y \equiv \tilde a^\mu (\psi_a) U_a(\psi_a) \partial_\mu \psi_a
\right)~.
\label{ex3}
\end{equation}
for which the first degeneracy condition (\ref{aVU}) is satisfied automatically as
\begin{equation}
A_{ia}^\mu
=
\tilde a^\mu(\psi_a) U_a(\psi_a) \frac{\partial \phi_i}{\partial Y}(\psi_a, Y)
\equiv 
a^\mu (\psi_a, Y)
U_a(\psi_a)
V_i(\psi_a, Y)
~.
\end{equation}
In this expression
$V_i$ 
is normalized as $V_i V_i=1$, that is,
\begin{equation}
V_i \equiv \frac{1}{v(\psi_a, Y)} \frac{\partial \phi_i}{\partial Y}(\psi_a, Y)~,
\quad
a^\mu(\psi_a, Y) \equiv v(\psi_a, Y)\tilde a^\mu(\psi_a)~,
\quad
V_iV_i = 1~,
\end{equation}
where $v(\psi_a, Y)$ is a normalization factor.
For simplicity, we assume $v\neq 0$ in the following.

Following the procedure explained in Appendix~\ref{App:Setting-U}, 
we set $U_a = (0,1)$, $m_a = (1,0)$ without loss of generality.
Then, using $n_i = v^{-1}\epsilon_{ij}\partial \phi_j / \partial Y$,
the second degeneracy condition (\ref{nBm}) is evaluated as
\begin{gather}
B_{ia}
=
\frac{\partial \phi_i}{\partial \psi_a}
+ \frac{\partial \phi_i}{\partial Y} \frac{\partial \tilde a^\mu}{\partial \psi_a}\partial_\mu \psi_2
=
\frac{\partial \phi_i}{\partial \psi_a}
+ v\,V_i
 \frac{\partial \tilde a^\mu}{\partial \psi_a}\partial_\mu \psi_2
~,
\\
n_i B_{ia} m_a
=
v^{-1}
\epsilon_{ij}
\frac{\partial \phi_i}{\partial \psi_1}
\frac{\partial \phi_j}{\partial Y}
= 0~.
\label{ex3_deg2}
\end{gather}
Equation (\ref{ex3_deg2}) implies that, provided $v\neq 0$,
\begin{equation}
\frac{\frac{\partial \phi_1}{\partial \psi_1}}{\frac{\partial \phi_1}{\partial Y}}
=
\frac{\frac{\partial \phi_2}{\partial \psi_1}}{\frac{\partial \phi_2}{\partial Y}}~,
\end{equation}
that is, gradient vectors $\bigl(\frac{\partial \phi_1}{\partial \psi_1}, \frac{\partial \phi_1}{\partial Y}\bigr)$ and $\bigl(\frac{\partial \phi_2}{\partial \psi_1}, \frac{\partial \phi_2}{\partial Y}\bigr)$ are parallel to each other in the $(\psi_1, Y)$ space. Then it follows that
\begin{equation}
\phi_2 = F\left(
\psi_2, \phi_1(\psi_a, Y)
\right)~,
\label{ex3_phi2}
\end{equation}
where $F(\psi_2, \phi_1)$ is an arbitrary function of $\phi_1$ and $\psi_2$.%
\footnote{When the $(\psi_1, Y)$ space is separated into connected sets by borders on which the gradient vector $\bigl(\frac{\partial \phi_1}{\partial \psi_1}, \frac{\partial \phi_1}{\partial Y}\bigr)$ vanishes, 
the function form of $F(\psi_2, \phi_1(\psi_a, Y))$  may be different on each connected sets on the $(\psi_1, Y)$ space. The non-degeneracy conditions however imply $\partial\phi_1/\partial Y\neq 0$, 
and it guarantees that $F(\psi_2, \phi_1)$ is given uniquely on the entire $(\psi_1, Y)$ space.}

Let us examine the non-degeneracy conditions (\ref{nBU}).
One of them is given by
\begin{align}
0 \neq n_i B_{ia} U_a
&= n_i B_{i2} 
= v^{-1} \epsilon_{ij} \frac{\partial \phi_i}{\partial \psi_2} \frac{\partial\phi_j}{\partial Y}
=
v^{-1}
\left(
  \frac{\partial \phi_1}{\partial \psi_2} \frac{\partial\phi_2}{\partial Y}
- \frac{\partial \phi_2}{\partial \psi_2} \frac{\partial\phi_1}{\partial Y}
\right)
\notag \\
&=
v^{-1}
\left[
  \frac{\partial \phi_1}{\partial \psi_2}
\frac{\partial F}{\partial \phi_1}\frac{\partial \phi_1}{\partial Y}
- 
\left(
\frac{\partial F}{\partial \psi_2}
+ \frac{\partial F}{\partial\phi_1} \frac{\partial \phi_1}{\partial \phi_2} 
\right)
\frac{\partial\phi_1}{\partial Y}
\right]
=
-v^{-1} \frac{\partial F}{\partial\phi_1}\frac{\partial\phi_1}{\partial Y}~.
\label{ex3_nondeg-1}
\end{align}
This equation implies that $v, \partial F/\partial \phi_1, \partial\phi_1/\partial Y$ must not vanish.
The other non-degeneracy condition is given by
\begin{align}
0&\neq 
\left(
V_i B_{ia} - a^\mu \partial_\mu U_a
\right) m_a
=
V_i \left(
\frac{\partial \phi_i}{\partial \psi_1}
+ v\,V_i
 \frac{\partial \tilde a^\mu}{\partial \psi_1}\partial_\mu \psi_2
\right)
=
V_i
\frac{\partial \phi_i}{\partial \psi_1}
+ v
 \frac{\partial \tilde a^\mu}{\partial \psi_1}\partial_\mu \psi_2~.
\label{ex3_nondeg-2}
\end{align}

To summarize, a transformation~(\ref{ex3}) is invertible if $\phi_2$ is given by Eq.~(\ref{ex3_phi2}) and Eqs.~(\ref{ex3_nondeg-1}), (\ref{ex3_nondeg-2}) are satisfied.

\section{No-go for disformal transformation of the metric with higher derivatives}
\label{sec:no-go}

Here we apply our approach to disformal metric transformation. 
The disformal transformations involving only one derivative of the scalar field $\chi$,
\begin{equation}
\label{disform1}
\tilde{g}_{\mu\nu} = C(\chi,X)g_{\mu\nu} + D(\chi,X) \nabla_\mu\chi \nabla_\nu\chi~,
\end{equation}
where $X=\partial_\mu\chi \partial^\mu\chi$, are invertible, provided that $C\left(C-\frac{dC}{dX} X - \frac{dD}{dX} X^2\right)\neq 0$, see e.g.~\cite{Zumalacarregui:2013pma,Domenech:2015tca}.
This follows from the fact that from the above expression one can straightforwardly express the metric $g_{\mu\nu}$ in terms of $\tilde{g}_{\mu\nu}$, $\chi$ and $\nabla_\mu\chi$.
Thus the transformation (\ref{disform1}) is one-to-one change of variables $(\chi,g_{\mu\nu}) \leftrightarrow  (\chi,\tilde{g}_{\mu\nu})$ as long as the above condition is satisfied. 
It is useful to check that the transformation (\ref{disform1}) is invertible by applying our method. Although the method we developed above does not apply metric transformations in general because a metric is a tensor and has more than 2 components except for one dimensional spacetime, we can use it when applied to particular ansatze that contain scalar functions only.
Indeed, let us restrict ourselves to the case of the homogeneous cosmology, 
\begin{equation}
\label{ansatzcosm}
g_{\mu\nu}dx^\mu dx^\nu  = - \mathfrak{n}(t) dt^2 + \mathfrak{a}(t) d{\bf x}^2, \quad \chi  =\chi(t),
\end{equation}
where $\mathfrak{a}(t)$ is the squared of the scale factor and $\mathfrak{n}(t)$ is the squared of the lapse function. Note that we keep $\mathfrak{n}(t)$ in the ansatz~(\ref{ansatzcosm}), since it changes under the transformation~(\ref{disform1}). For~(\ref{ansatzcosm}) we obtain under the disformal transformation~(\ref{disform1}),
\begin{equation}
    \label{transformna1}
    \tilde{\mathfrak{n}} =C\mathfrak{n} - D\dot\chi^2, \quad
    \tilde{\mathfrak{a}} = C\mathfrak{a},
\end{equation}
where a dot denotes a $t$ derivative, and
\begin{equation}
C
= C\left(\chi,X\right)
= C\left(\chi(t), - \frac{\dot\chi^2(t)}{\mathfrak{n}(t) } \right),
\qquad
D
= D\left(\chi(t), - \frac{\dot\chi^2(t)}{\mathfrak{n}(t) } \right).
\end{equation}
Note that the above transformation can be considered as the change of variables $\{\mathfrak{n},\mathfrak{a}\to \tilde{\mathfrak{n}},\tilde{\mathfrak{a}}\}$ (with 
the time-dependent external function $\chi(t)$), that does not involve derivatives. Therefore, by the virtue of the standard theorem on invertibility, the transformation~(\ref{transformna1}) is invertible if $\det B_{ia}\neq 0$, where $B_{ia}$ is given by,
\begin{equation}
\renewcommand*{\arraystretch}{1.4}
B_{ia}=
\begin{pmatrix}
C - X C_X - X^2 D_X & 0
\\
- \frac{\mathfrak{a}}{\mathfrak{n}} X C_X &  C \\
\end{pmatrix}, 
\end{equation}
so that we obtain $C\left(C-\frac{dC}{dX} X - \frac{dD}{dX} X^2\right)\neq 0$, which reproduces the result we cited above. 

On the other hand, one can consider a more general disformal transformation by including two derivatives of the scalar as follows,  
\begin{equation}
\label{disform2}
\tilde{g}_{\mu\nu} = C(\chi,X)g_{\mu\nu} + D(\chi,X) \nabla_\mu\chi \nabla_\nu\chi+ E(\chi,X) \nabla_\mu\nabla_\nu\chi.
\end{equation}
For the above transformation one cannot directly express $g_{\mu\nu}$ in terms of $\tilde{g}_{\mu\nu}$, since the last term of the r.h.s.\ of~(\ref{disform2}) also contains the metric $g_{\mu\nu}$.
Therefore, it has been conjectured that the inverse transformation of~(\ref{disform2}) does not exist, see Ref.~\cite{Ezquiaga:2017ner}.
However, to the best of our knowledge, this has not yet been proven.
Indeed, although the simple inverse of~(\ref{disform2}) does not exist, it does not necessarily mean that there is no more complicated inverse transformation. 
Using our method, however, we are able to demonstrate that the transformation~(\ref{disform2}) with non-zero $E$ is not invertible indeed. 

To do this, let us assume that the transformation (\ref{disform2}) is invertible, and we will see that this assumption leads to a contradiction.
The FRW  homogeneous ansatz (\ref{FRW}) expresses a subspace of functional space described with $g_{\mu\nu}$ and $\chi$. 
The invertibility limited to this subspace gives the necessary condition of invertibility for full functional space. 
Here, we will show the violation of this necessary condition, which leads to a contradiction.
From~(\ref{disform2}) and~(\ref{ansatzcosm}) we have,
\begin{equation}
\label{transformNa}
\tilde{\mathfrak{n}} = 
C  \mathfrak{n}- D\dot\chi^2 - E\left(\ddot{\chi} -\frac{\dot\chi\dot{\mathfrak{n}}}{2\mathfrak{n}}\right)
,\quad
\tilde{\mathfrak{a}} = C \mathfrak{a} -E \frac{\dot{\chi}
\dot{\mathfrak{a}}}{2\mathfrak{n}},
\end{equation}
which is a generalization of~(\ref{transformna1}) for the case of non-zero $E$.

As in the case considered above,  
we treat~(\ref{transformNa}) as a transformation relating a set of two variables $\bigl(\mathfrak{n}(t), \mathfrak{a}(t)\bigr)$ with $\bigl(\tilde{\mathfrak{n}}(t), \tilde{\mathfrak{a}}(t)\bigr)$ with 
the time-dependent external function $\chi(t)$. Contrary to standard disformal transformation, in this case the transformation involves first derivatives of the variables $\mathfrak{n}(t)$ and $\mathfrak{a}(t)$, therefore we should apply our method to check whether the conditions for this transformation to be invertible are satisfied.
We have,
\begin{equation}
\label{disformalA}
A_{ia} 
= \frac{E\dot\chi}{2\mathfrak{n}} \begin{pmatrix}
1 &0\\
0& -1 \\
\end{pmatrix}.
\end{equation}
According to our discussion above, the leading-order condition for the transformation given by (\ref{disform2}) to be invertible is the vanishing of the determinant of~(\ref{disformalA}). As we can see, this condition is clearly violated for non-zero $E$, therefore we arrive to the conclusion that the transformation~(\ref{disform2}) is not invertible unless $E=0$. 

Once we generalize the structure of the higher-derivative term as~\cite{Minamitsuji:2021dkf,Alinea:2020laa}
\begin{equation}
	\tilde{g}_{\mu\nu} = \mathcal{F}_0 g_{\mu\nu}+ \mathcal{F}_1 \chi_\mu\chi_\nu+ \mathcal{F}_2 \chi_{\mu\nu} +  \mathcal{F}_3 \chi_{\left(\mu\right.}X_{\left.\nu\right)} 
	+ \mathcal{F}_4 X_\mu X_\nu + \mathcal{F}_5 \chi_\mu^\alpha\chi_{\nu \alpha},
\end{equation}
where $\mathcal{F}_i = \mathcal{F}_i(\phi,X,\mathcal{B},\mathcal{Y}, \mathcal{Z}, \mathcal{W})$ and
\begin{equation}
\mathcal{B} = \nabla_\mu\nabla^\mu\chi,
\quad
\mathcal{Y} = \nabla_\mu \chi  \nabla^\mu X,
\quad
\mathcal{Z} = \nabla_\mu X \nabla^\mu X,
\quad
\mathcal{W}= \nabla^\mu\nabla^\nu\chi \nabla_\mu\nabla_\nu\chi,
\end{equation}
the structure of $A_{ia}$ is changed and the no-go result obtained above for the transformation~(\ref{disform2}) could be evaded. We examine such a possibility in Appendix~\ref{app:gendisformal}.

\section{Discussions}
\label{sec:discussions}

In this work we focused on field transformations that involve up to first-order derivative of fields, $\phi_i = \bar \phi_i(\psi_a, \partial_\alpha \psi_a, x^\mu)$, between two fields $\psi_a$ and other two fields $\phi_i$, and showed conditions for this transformation to be invertible.
A field transformation of this type changes the number of derivatives
acting on the fields, hence in general it changes number of degree of freedom.
When the transformation function satisfies certain conditions, however, appearance of additional degrees of freedom is hindered and then the transformation can be invertible.

We emphasize that the degeneracy conditions and the procedure to derive the complete set of the invertibility conditions given in section~\ref{sec:necessary_general} apply to field transformations that involve arbitrary number of fields and arbitrary-order  derivatives, though we presented expressions only for transformations with first derivatives just for simplicity.
In the following part of this work,
for simplicity, we limited our scope to the simplest case that the transformation maps two fields $\phi_i~(i=1,2)$ to other two fields $\psi_a~(a=1,2)$, and then derived the conditions for this transformation to be invertible. This is because, as we have shown, there is no invertible transformation with derivatives for one field case and hence two field case is the simplest case.
To derive the necessary conditions for the invertibility, we employed the method of characteristics for a partial differential equations in section~\ref{sec:necessary}.
If a transformation is invertible, the number of characteristic surfaces, which corresponds to the number of physical degrees of freedom, must be invariant. 
The derivatives contained in the transformation generates extra characteristic surfaces in general, and then
the necessary conditions are obtained by demanding that extra characteristic surfaces are removed so that the total number of characteristic surfaces is invariant.
After deriving the necessary conditions, in section~\ref{Sec:Suf} we confirmed that they are actually sufficient.
It turned out that the invertibility conditions are composed of two degeneracy conditions (\ref{aVU}), (\ref{nBm}) and two non-degeneracy conditions given in Eq.~(\ref{nBU}).

As an application of the thus derived invertibility conditions, in section~\ref{sec:examples} we showed some examples of invertible transformations satisfying the invertibility conditions.
The invertibility conditions can be regarded as equations for the function of the field transformation, and if we could construct their general solution we could obtain the most-general invertible transformation.
Instead of finding the general solution, we proposed some ansatz for the transformation for which a part of the invertibility conditions are automatically satisfied, and as a result we obtained three kinds of invertible transformations as non-trivial examples. Although they may not be most general, they span broad class of invertible transformations and would provide a basis for construction and classification of various invertible transformations.

As another application, in section~\ref{sec:no-go} we considered a higher-derivative extension of the disformal transformation in gravity and examined its invertibility. Using our invertibility conditions, just by simple calculations we have explicitly shown that a disformal-type transformation associated with second derivative of the scalar field cannot be invertible, which is the first rigorous proof as far as we are aware of.

Several directions of the future research are in order.
In this work we considered the simplest case that the transformation involves only two scalar fields up to their first derivatives.
Provided that our method can be generalized to transformations involving both a scalar field and a metric,
we will be able to apply our results to studies on scalar-tensor theories and various modified gravity theories.
For example, invertible disformal transformations were utilized to generate and classify the so-called DHOST theories from a simpler theory, the Horndeski theory.
This scheme may be generalized to incorporate higher-order derivatives if we could generalize the disformal transformation by introducing higher derivatives.
Such an applications to modified gravity theory will be an ultimate goal of this study.
As a first step for such a goal, it would be useful to consider a generalization to involve more than two scalar fields and more than first-order derivatives. Such generalizations within transformations of scalar fields, and also further generalizations including more fields such as metric will be discussed in our future work.

\section*{Acknowledgments}

K.~I.\ is supported by JSPS Grants-in-Aid for  Scientific Research
(A) (No.\ JP17H01091) and Scientific Research (B) (No.\ JP20H01902).
N.~T.\ is supported in part by JSPS Grant-in-Aid for Scientific Research No.\ JP18K03623.
M.~Y.\ is supported in part by JSPS Grant-in-Aid for Scientific Research Numbers JP18K18764, JP21H01080, JP21H00069.

\appendix

\section{Non-invertibility of field-number changing transformations}
\label{App:field-number}

Let us 
consider a transformation $\phi_i=\phi_i(\psi_1, \partial\psi_1)  ~(i=1,2)$, for which the number of fields decreases but the number of derivative increases.
Naively thinking, this transformation preserves the degree of freedom because the product of the number of fields and the order of derivatives is invariant under this transformation. 
To examine this expectation, let us evaluate the invertibility conditions.
For simplicity we work in the one-dimensional case where the fields depend only on one variable, that is, we work in a point particle system.
The first degeneracy condition (\ref{aVU}) is given by
\begin{equation}
A_{ia} = \frac{\partial \phi_i}{\partial \dot \psi_a}
=
\begin{pmatrix}
\frac{\partial \phi_1}{\partial\dot\psi_1} & 0 \\
\frac{\partial \phi_2}{\partial\dot\psi_1} & 0 
\end{pmatrix}
=a V_i U_a~,
\qquad
V_i = \frac1a \frac{\partial \phi_i}{\partial\dot\psi_1}~,
\qquad
U_a = (1,0)~,
\end{equation}
where $a$ is chosen to normalize $V_i$ as $V_i V_i = 1$.
Then, using $n_i = \epsilon_{ij} V_j$ and $m_a = \epsilon_{ab}U_b = (0,-1)$, the second degeneracy condition (\ref{nBm}) and the non-degeneracy conditions (\ref{nBU}) are given by
\begin{equation}
B_{ia} = \frac{\partial\phi_i}{\partial \psi_a}
=
\begin{pmatrix}
\frac{\partial \phi_1}{\partial\psi_1} & 0 \\
\frac{\partial \phi_2}{\partial\psi_1} & 0 
\end{pmatrix}
=
\tilde V_i U_a
~,
\qquad
\tilde V_i = \frac{\partial \phi_i}{\partial\psi_1}~,
\end{equation}
\begin{equation}
n_i B_{ia} m_a = 0,
\qquad
n_i B_{ia} U_a = n_i \tilde V_i ~,
\qquad
\left( V_i B_{ia} - a \dot U_a \right) m_a = 0~.
\end{equation}
While $n_i B_{ia} U_a = n_i \tilde V_i$ may not hold, $\bigl( V_i B_{ia} - a \dot U_a \bigr) m_a$ vanishes identically, hence the non-degeneracy condition among the invertibility conditions is violated. Then we can judge that the transformation $\phi_i=\phi_i(\psi_1, \partial\psi_1)  ~(i=1,2)$ is not invertible.
This result can be understood as follows. If this transformation were invertible, then an inverse transformation $\psi_1 = f_1(\phi_1, \phi_2), ~ \dot \psi_1 = f_2(\phi_1, \phi_2)$ would exist.
This implies that there exists a constraint $\dot f_1 = f_2$ between $\phi_1$ and $\phi_2$. This is in contradiction with the fact that $\phi_1$ and $\phi_2$ are independent variables, then it implies that the transformation cannot be invertible.

\section{Derivation of Eq.~(\ref{eq:C})}
\label{App:derivingformula}

In this section we sketch the derivation of the formula (\ref{eq:C}), based on which the necessary conditions for the invertibility are derived.
Acting $\partial_{\mu_n}$ on $\phi_i = \bar \phi_i(\psi_a, \partial_\alpha \psi_a, x^\mu)$ gives
\begin{equation}
\partial_{\mu_n}\phi_i(\psi_a, \partial_\alpha \psi_a, x^\mu)
=
\frac{\partial\bar\phi_i}{\partial \psi_a}
\partial_{\mu_n}\psi_a
+ 
\frac{\partial\bar\phi_i}{\partial \left(\partial_\alpha \psi_a\right)}
\partial_{\mu_n}\partial_\alpha\psi_a
+\frac{\partial \bar\phi}{\partial x^{\mu_n}}~,
\label{delphi}
\end{equation}
where the third term is the partial derivative with respect to the explicit $x^\mu$ dependence of $\bar\phi_i$.
Then, the highest derivative term of $\psi_a$ contained in $\partial_{\mu_1\ldots \mu_n}\phi_i$ is generated when all the derivatives other than $\partial_{\mu_n}$ act on $\partial_{\mu_n}\partial_\alpha\psi_a$ in Eq.~(\ref{delphi}), that is,
\begin{equation}
\partial_{\mu_1\ldots \mu_n}\phi_i
\ni
\frac{\partial\bar\phi_i}{\partial \left(\partial_\alpha \psi_a\right)}
\partial_{\mu_1\ldots\mu_n\alpha}\psi_a
=
A_{ia}^\alpha \partial_{\mu_1\ldots\mu_n\alpha}\psi_a~.
\end{equation}
This term is the origin of the leading $\partial^{n+1}\psi_a$ term of Eq.~(\ref{eq:C}).
The subleading $\partial^n\psi_a$ term is composed of the following two contributions.
The first one is generated when $\partial_{\mu_1\ldots \mu_{n-1}}$ acts on $\partial_{\mu_n}\psi_a$ in Eq.~(\ref{delphi}), that is,
\begin{equation}
\partial_{\mu_1\ldots \mu_n}\phi_i
\ni
\frac{\partial\bar\phi_i}{\partial \psi_a}
\partial_{\mu_1\ldots \mu_n}\psi_a
=
B_{ia} \partial_{\mu_1\ldots \mu_n}\psi_a~.
\end{equation}
The second one is generated when $n-1$ derivatives among $\partial_{\mu_1\ldots \mu_n}$ are consumed to generate $A^{\mu_n}_{ia} \partial^n\psi_a$ and the other one derivative acts directly on $\phi_i$, that is,
\begin{equation}
\partial_{\mu_1\ldots \mu_n}\phi_i
\ni
\sum_{k=1}^n
\frac{\partial^2\bar\phi_i}{\partial x^{\mu_k} \partial \left(\partial_\alpha \psi_a\right)}
\partial_{\mu_1\ldots\mu_{k-1}\mu_{k+1}\ldots \mu_n\alpha}\psi_a
=
\sum_{k=1}^n
\partial_{\mu_k} A^\alpha_{ia}
\partial_{\mu_1\ldots\mu_{k-1}\mu_{k+1}\ldots \mu_n\alpha}\psi_a~.
\end{equation}
Contracting with a totally-symmetric coefficient $K^{(\mu_1\ldots \mu_n)}_{bi}$, 
the expression appearing in Eq.~(\ref{eq:C}) is obtained as
\begin{equation}
K^{(\mu_1\ldots \mu_n)}_{bi}
\partial_{\mu_1\ldots \mu_n}\phi_i
\ni
K^{(\mu_1\ldots \mu_n)}_{bi}
\sum_{k=1}^n
\partial_{\mu_k} A^\alpha_{ia}
\partial_{\mu_1\ldots\mu_{k-1}\mu_{k+1}\ldots \mu_n\alpha}\psi_a
=
n \, K^{(\alpha \mu_1\ldots \mu_{n-1})}_{bi}
\partial_{\alpha} A^{\mu_n}_{ia}
\partial_{\mu_1\ldots \mu_n}\psi_a~.
\end{equation}
The $\partial^{n-k}\psi_a$ term of Eq.~(\ref{eq:C}) is obtained in a similar manner.
The $\partial^k B_{ia}$ term is obtained by using $n-k$ derivatives to generate $\partial^{n-k}\psi_a$ and by acting the other derivatives on $\phi_i$ directly.
The $\partial^{k+1}A^\mu_{ia}$ term is obtained by using $n-k-1$ derivatives to generate $\partial^{n-k}\psi_a$ and by acting the other derivatives on $\phi_i$ directly. The coefficients of each term, $\binom{n}{k}$ and $\binom{n}{k+1}$, correspond to the number of combinations of the derivatives among $\partial^n$ used to generate $\partial^k \psi_a$.

\section{Derivation of Eq.~(\ref{subsubleadingeq_2field})}
\label{App:longcalculation}

We give the derivation of Eq.~(\ref{subsubleadingeq_2field}) based on Eq.~(\ref{eq:C}) and using the degeneracy conditions in the two-field case, Eqs.~(\ref{detA}) and (\ref{2sub}).
We also use results given in section~\ref{sec:two-fields}.

We start the derivation from 
\begin{align}
 &\dbtilde K^{(\mu_1\dots\mu_n)} \bar A^{\mu_{n+1}}_{bi} \bar A^{2,\mu_{n+2}\mu_{n+3}} \partial_{\mu_1\dots \mu_{n+3}} \phi_i 
\nonumber \\ &
-
\left[
\dbtilde K^{(\mu_1\dots\mu_n)} {\cal A}^{\mu_{n+1}}_{2,bc} \tilde A^{\mu_{n+2}}_{ci}
-
\left(
n\dbtilde K^{(\alpha\mu_1\dots\mu_{n-1})}\bar A^{2,\mu_{n}\mu_{n+1}}
+2 \dbtilde K^{(\mu_1\dots\mu_{n})} \bar A^{2,\mu_{n+1}\alpha}
\right)
\partial_\alpha \bar A^{\mu_{n+2}}_{bi} 
\right]
\partial_{\mu_1\dots \mu_{n+2}} \phi_i,
\label{difference}
\end{align}
which is given by $\partial^{n+2}\psi$ terms and lower-order terms once rewritten in terms of $\psi$ using Eq.~(\ref{eq:C}), as explained in section~\ref{sec:two-fields}.

Using Eq.~(\ref{eq:C}), the first term of Eq.~(\ref{difference}) is expressed in terms of $\psi$ as
\begin{align}
&\dbtilde K^{(\mu_1\dots\mu_n)} \bar A^{\mu_{n+1}}_{bi} \bar A^{2,\mu_{n+2}\mu_{n+3}} \partial_{\mu_1\dots \mu_{n+3}} \phi_i
\notag \\
&=\left[\text{$\partial^{n+3}\psi$ term}\right]
\nonumber\\& 
+\Biggl\{\biggl[
\frac{n(n-1)}{2}\dbtilde K^{(\alpha_1\alpha_2 \mu_1\dots\mu_{n-2})} \bar A^{\mu_{n-1}}_{bi} \bar A^{2,\mu_{n}\mu_{n+1}}
+n\,\dbtilde K^{(\alpha_1 \mu_1\dots\mu_{n-1})} \bar A^{\alpha_2}_{bi} \bar A^{2,\mu_{n}\mu_{n+1}}
\nonumber \\ & \qquad\quad
+2n\,\dbtilde K^{(\alpha_1 \mu_1\dots\mu_{n-1})} \bar A^{\mu_{n}}_{bi} \bar A^{2,\mu_{n+1}\alpha_2}
+\dbtilde K^{(\mu_1\dots\mu_n)}
\left(
2 \bar A^{\alpha_1}_{bi} \bar A^{2,\mu_{n+1}\alpha_2}
+ \bar A^{\mu_{n+1}}_{bi} \bar A^{2,\alpha_1\alpha_2}
\right)
\biggr]
\partial_{\alpha_1\alpha_2}A^{\mu_{n+2}}_{ia}
\notag \\
&\quad
+ 
\left[
n\, \dbtilde K^{(\alpha\mu_1\dots\mu_{n-1})} \bar A^{\mu_{n}}_{bi} \bar A^{2,\mu_{n+1}\mu_{n+2}} 
+\dbtilde K^{(\mu_1\dots\mu_n)} 
\left(
\bar A^{\alpha}_{bi} \bar A^{2,\mu_{n+1}\mu_{n+2}}  
+2 \bar A^{\mu_{n+1}}_{bi} \bar A^{2,\mu_{n+2}\alpha} \right) \right]
\partial_\alpha B_{ia}
\Biggr\} \partial_{\mu_1\ldots \mu_{n+2}} \psi_a
\notag \\
&
+\cO{\partial^{n+1} \psi_a} ,
\label{firsttermexpanded}
\end{align}
where the terms appearing on the right-hand side are classified according to the positions of the indices $\alpha, \alpha_1, \alpha_2$.
Likewise, the second term of Eq.~(\ref{difference}) is expressed as
\begin{align}
&\left[
\dbtilde K^{(\mu_1\dots\mu_n)} {\cal A}^{\mu_{n+1}}_{2,bc} \tilde A^{\mu_{n+2}}_{ci}
-
\left(
n\dbtilde K^{(\alpha\mu_1\dots\mu_{n-1})}\bar A^{2,\mu_{n}\mu_{n+1}}
+2 \dbtilde K^{(\mu_1\dots\mu_{n})} \bar A^{2,\mu_{n+1}\alpha}
\right)
\partial_\alpha \bar A^{\mu_{n+2}}_{bi} 
\right]
\partial_{\mu_1\dots \mu_{n+2}} \phi_i
\notag \\
&=
\left[\text{$\partial^{n+3}\psi$ term}\right]
\notag \\ &
+\Biggl\{
\left[
\dbtilde K^{(\mu_1\dots\mu_n)} {\cal A}^{\mu_{n+1}}_{2,bc} \tilde A^{\mu_{n+2}}_{ci}
-
\left(
n\dbtilde K^{(\alpha\mu_1\dots\mu_{n-1})}\bar A^{2,\mu_{n}\mu_{n+1}}
+2 \dbtilde K^{(\mu_1\dots\mu_{n})} \bar A^{2,\mu_{n+1}\alpha}
\right)
\partial_\alpha \bar A^{\mu_{n+2}}_{bi} 
\right]B_{ia}
\nonumber \\ & \quad
+\Bigl\{
n\, \dbtilde K^{(\alpha_2\mu_1\dots\mu_{n-1})} {\cal A}^{\mu_{n}}_{2,bc} \tilde A^{\mu_{n+1}}_{ci}
+ 2\dbtilde K^{(\mu_1\dots\mu_n)} 
{\cal A}^{(\alpha_2}_{2,bc} \tilde A^{\mu_{n+1})}_{ci}
\nonumber \\ & \qquad
-n\Bigl[
(n-1)\dbtilde K^{(\alpha_1\alpha_2\mu_1\dots\mu_{n-2})} \bar A^{2,\mu_{n-1}\mu_{n}}\partial_{\alpha_1} \bar A^{\mu_{n+1}}_{bi} 
+\dbtilde K^{(\alpha_1\mu_1\dots\mu_{n-1})} 
\left(
2 \bar A^{2,\alpha_2\mu_{n}}\partial_{\alpha_1} \bar A^{\mu_{n+1}}_{bi}
+ \bar A^{2,\mu_{n}\mu_{n+1}} \partial_{\alpha_1} \bar A^{\alpha_2}_{bi}
\right)
\Bigr]
\nonumber \\ & \qquad
-2\Bigl(
n \,\dbtilde K^{(\alpha_2\mu_1\dots\mu_{n-1})} \bar A^{2,\mu_{n}\alpha_1} \partial_{\alpha_1} \bar A^{\mu_{n+1}}_{bi}
+ 2\dbtilde K^{(\mu_1\dots\mu_{n})}
\bar A^{2,\alpha_1(\alpha_2} \partial_{\alpha_1} \bar A^{\mu_{n+1})}_{bi}
\Bigr)
\Bigr\} \partial_{\alpha_2}A^{\mu_{n+2}}_{ia}
\Biggr\}
\partial_{\mu_1\ldots\mu_{n+2}} \psi_a
\notag \\
&
+\cO{\partial^{n+1} \psi_a} ,
\label{secondtermexpanded}
\end{align}
where the $\partial^{n+3}\psi$ term on the right-hand side is the same as that in Eq.~(\ref{firsttermexpanded}).
Then, in Eq.~(\ref{difference}) the $\partial^{n+3}\psi$ terms cancel out, and the remaining terms are given by the difference between Eqs.~(\ref{firsttermexpanded}) and (\ref{secondtermexpanded}). Reorganizing the terms according to their tensor structures, it can be expressed as
\begin{align}
 &\dbtilde K^{(\mu_1\dots\mu_n)} \bar A^{\mu_{n+1}}_{bi} \bar A^{2,\mu_{n+2}\mu_{n+3}} \partial_{\mu_1\dots \mu_{n+3}} \phi_i 
\nonumber \\ &
-
\left[
\dbtilde K^{(\mu_1\dots\mu_n)} {\cal A}^{\mu_{n+1}}_{2,bc} \tilde A^{\mu_{n+2}}_{ci}
-
\left(
n\dbtilde K^{(\alpha\mu_1\dots\mu_{n-1})}\bar A^{2,\mu_{n}\mu_{n+1}}
+2 \dbtilde K^{(\mu_1\dots\mu_{n})} \bar A^{2,\mu_{n+1}\alpha}
\right)
\partial_\alpha \bar A^{\mu_{n+2}}_{bi} 
\right]
\partial_{\mu_1\dots \mu_{n+2}} \phi_i
\notag \\
&= \Bigg\{
\frac{n(n-1)}{2} \dbtilde K^{(\alpha_1 \alpha_2 \mu_1\dots\mu_{n-2})}  \bar A^{2,\mu_{n-1}\mu_n} 
\left[ \bar A^{\mu_{n+1}}_{bi} \partial _{\alpha_1 \alpha_2} A^{\mu_{n+2}}_{ia} 
 +2 \left(\partial _{\alpha_1} \bar A^{\mu_{n+1}}_{bi}\right)  \partial_{\alpha_2} A^{\mu_{n+2}}_{ia} \right]
\nonumber \\ & \quad
+n\, \dbtilde K^{(\alpha_1 \mu_1\dots\mu_{n-1})} \Bigl\{ 
-{\cal A}^{\mu_{n}}_{2,bc} \tilde A^{\mu_{n+1}}_{ci}  \partial_{\alpha_1} A^{\mu_{n+2}}_{ia}
\nonumber \\ & \hspace{32mm}
+2\bar A^{2,\mu_n\alpha_2} \left[
\bar A^{\mu_{n+1}}_{bi} \partial_{\alpha_1\alpha_2} A^{\mu_{n+2}}_{ia}
+\left( \partial_{\alpha_1} \bar A^{\mu_{n+1}}_{bi}\right) \partial_{\alpha_2} A^{\mu_{n+2}}_{ia} 
+\left( \partial_{\alpha_2} \bar A^{\mu_{n+1}}_{bi}\right) \partial_{\alpha_1} A^{\mu_{n+2}}_{ia} 
 \right]
 \nonumber \\ & \hspace{32mm}
+\bar A^{2,\mu_{n}\mu_{n+1}} 
\left[ \bar A^{\mu_{n+2}}_{bi} \partial_{\alpha_1} B_{ia}
+ \left( \partial_{\alpha_1}  \bar A^{\mu_{n+2}}_{bi}  \right) B_{ia}
+ \bar A^{\alpha_2}_{bi} \partial_{\alpha_1\alpha_2}A^{\mu_{n+2}}_{ia}
+ \left(\partial_{\alpha_1} \bar A^{\alpha_2}_{bi}\right)\partial_{\alpha_2}A^{\mu_{n+2}}_{ia}
\right]
\Bigr\}
\nonumber \\ & \quad
+\dbtilde K^{(\mu_1\dots\mu_{n})} \Bigl[
\bar A^{\alpha_1}_{bi} \bar A^{2,\mu_{n+1}\mu_{n+2}}\partial_{\alpha_1} B_{ia}
-2{\cal A}^{(\alpha_2}_{2,bc} \tilde A^{\mu_{n+1})}_{ci} \partial_{\alpha_2} A^{\mu_{n+2}}_{ia}
-{\cal A}^{\mu_{n+1}}_{2,bc} \tilde A^{\mu_{n+2}}_{ci}B_{ia} 
\nonumber \\
&\hspace{23mm}
+2  \bar A^{2,\mu_{n+1}\alpha_1}
\left(
\partial_{\alpha_1} \bar A^{\mu_{n+2}}_{bi} 
B_{ia}
+\bar A^{\mu_{n+2}}_{bi} 
\partial_{\alpha_1} B_{ia} 
+\bar A^{\alpha_2}_{bi} 
\partial_{\alpha_1\alpha_2}A^{\mu_{n+2}}_{ia}
+  
\partial_{\alpha_1} \bar A^{\alpha_2}_{bi} \partial_{\alpha_2} A^{\mu_{n+2}}_{ia}
\right)
\nonumber \\
&\hspace{23mm}
+\bar A^{2,\alpha_1\alpha_2}\left(
\bar A^{\mu_{n+1}}_{bi} 
\partial_{\alpha_1\alpha_2}A^{\mu_{n+2}}_{ia}
+ 2\partial_{\alpha_1} \bar A^{\mu_{n+1}}_{bi} \partial_{\alpha_2} A^{\mu_{n+2}}_{ia}
\right)
\Bigr]
\Biggr\}
\partial_{\mu_1\dots \mu_{n+2}} \psi_a
+\cO{\partial^{n+1} \psi_a}
\notag \\
&= \bigg\{ 
-\frac{n(n-1)}{2} \dbtilde K^{(\alpha_1\alpha_2\mu_1\dots\mu_{n-2})}  \bar A^{2,\mu_{n-1}\mu_n}
\left( \partial _{\alpha_1 \alpha_2} \bar A^{\mu_{n+1}}_{bi} \right) A^{\mu_{n+2}}_{ia}
\nonumber \\ & \quad
+n\,\dbtilde K^{(\alpha_1 \mu_1\dots\mu_{n-1})} \left[
-{\cal A}^{\mu_{n}}_{2,bc} \tilde A^{\mu_{n+1}}_{ci}  \partial_{\alpha_1} A^{\mu_{n+2}}_{ia} 
-2\bar A^{2,\mu_n\alpha_2} \left(\partial_{\alpha_1\alpha_2} \bar A^{\mu_{n+1}}_{bi}\right)A^{\mu_{n+2}}_{ia}
+\bar A^{2,\mu_{n}\mu_{n+1}} \partial_{\alpha_1} {\cal A}^{\mu_{n+2}}_{ba} 
\right]
\nonumber \\ & \quad
+\dbtilde K^{(\mu_1\dots\mu_{n})}
\Bigl[
\bar A^{2,\mu_{n+1}\mu_{n+2}}  \bar A^{\alpha}_{bi}\partial_{\alpha} B_{ia}
 -2 {\cal A}^{(\alpha}_{2,bc} \tilde A^{\mu_{n+1})}_{ci}
\partial_{\alpha} A^{\mu_{n+2}}_{ia} 
-{\cal A}^{\mu_{n+1}}_{2,bc} \tilde A^{\mu_{n+2}}_{ci}B_{ia}
+2\bar A^{2,\mu_{n+1}\alpha} \partial_{\alpha} {\cal A}^{\mu_{n+2}}_{2,ba}
\nonumber \\ & \hspace{55mm}
- \bar A^{2,\alpha_1\alpha_2} \left( \partial_{\alpha_1\alpha_2}\bar A^{\mu_{n+1}}_{bi} \right)  A^{\mu_{n+2}}_{ia}
\Bigr]
\biggr\} \partial_{\mu_1\dots \mu_{n+2}} \psi_a
+\cO{\partial^{n+1} \psi_a}.
\label{longexpression}
\end{align}
At the final equality, we used Eq.~(\ref{calAdef}) and the identity that follows from the second derivative of Eq.~(\ref{AA}) in the two-field case, which is given by
\begin{equation}
\partial_\mu \partial_\nu \left(\bar A_{bi}^{(\alpha} A_{ia}^{\beta)}\right) 
=
\left(\partial_\mu \partial_\nu \bar A_{bi}^{(\alpha} \right) A_{ia}^{\beta)}
+\left(\partial_\mu \bar A_{bi}^{(\alpha} \right)\partial_\nu  A_{ia}^{\beta)}
+\left(\partial_\nu \bar A_{bi}^{(\alpha} \right)\partial_\mu  A_{ia}^{\beta)}
+ \bar A_{bi}^{(\alpha} \left(\partial_\mu \partial_\nu A_{ia}^{\beta)}\right)
=0.
\end{equation}

As argued in section~\ref{sec:two-fields}, to find the last condition for the invertibility, we should focus on the coefficient of the highest derivative term in the $\psi^\perp$ space, and this coefficient is obtained by replacing 
$\partial_{\mu_1\dots \mu_{n+2}} \psi_a$ by 
$\bar A^{\mu_3}_{aj}$ and symmetrizing over $\mu_1, \mu_2, \mu_3$ in Eq.~(\ref{longexpression}).
Below, let us evaluate each term in Eq.~(\ref{longexpression}) after this replacement.
First, the $\dbtilde K^{(\alpha_1\alpha_2\mu_1\dots\mu_{n-2})}$ term vanishes because it contains
$A^{(\mu_{n+2}}_{ia}\bar A^{\mu_3)}_{aj}=0$, which is enforced by Eq.~(\ref{calAdef}).
Next, the $\dbtilde K^{(\alpha_1 \mu_1\dots\mu_{n-1})}$ term also vanishes as follows.
\begin{align}
&\left[
-{\cal A}^{(\mu_{n}}_{2,bc} \tilde A^{\mu_{n+1}}_{ci}  \partial_{\alpha_1} A^{\mu_{n+2}}_{ia} 
-2\bar A^{2,\alpha_2(\mu_n} \left(\partial_{\alpha_1\alpha_2} \bar A^{\mu_{n+1}}_{bi}\right)A^{\mu_{n+2}}_{ia}
+\bar A^{2,(\mu_{n}\mu_{n+1}} \partial_{\alpha_1} {\cal A}^{\mu_{n+2}}_{ba} 
\right]\bar A^{\mu_3)}_{aj}
\notag \\
&=
\left(
-{\cal A}^{(\mu_{n}}_{2,bc} \tilde A^{\mu_{n+1}}_{ci}  \partial_{\alpha_1} A^{\mu_{n+2}}_{ia} 
+\bar A^{2,(\mu_{n}\mu_{n+1}} \partial_{\alpha_1} {\cal A}^{\mu_{n+2}}_{ba} 
\right)
\bar A^{\mu_3)}_{aj}
\notag \\
&=
\left(
{\cal A}^{(\mu_{n}}_{2,bc} \tilde A^{\mu_{n+1}}_{ci}   A^{\mu_{n+2}}_{ia} 
-\bar A^{2,(\mu_{n}\mu_{n+1}}  {\cal A}^{\mu_{n+2}}_{ba} 
\right)
\partial_{\alpha_1}\bar A^{\mu_3)}_{aj}
\notag \\
&=
\left[
{\cal A}^{(\mu_{n}}_{2,bc} \tilde A^{\mu_{n+1}}_{ci}   A^{\mu_{n+2}}_{ia} 
-
{\cal A}^{(\mu_n}_{bc}
\left(
{\bar A}_{ci}^{\mu_{n+1}} {\bar A}_{ai}^{\mu_{n+2}}
+
\tilde A^{\mu_{n+1}}_{ci} A^{\mu_{n+2}}_{ia}
\right)
\right]
\partial_{\alpha_1}\bar A^{\mu_3)}_{aj}=0.
\end{align}
The first equality follows from Eq.~(\ref{calAdef}).
The second equality follows from Eq.~(\ref{AbardA=0}) and the first derivative of Eq.~(\ref{2sub}), that is,
\begin{equation}
\partial_{\alpha_1} {\cal A}^{(\mu_{n+2}}_{ba} \bar A^{\mu_3)}_{aj}
=
- {\cal A}^{(\mu_{n+2}}_{ba} \partial_{\alpha_1} \bar A^{\mu_3)}_{aj}.
\end{equation}
At the third equality we replaced $\bar A^{2,\mu_{n}\mu_{n+1}}$ using Eq.~(\ref{project_2field}).
Then this equation can be shown to vanish using Eq.~(\ref{2sub}).
Lastly, in the $\dbtilde K^{(\mu_1\dots\mu_{n})}$ term of Eq.~(\ref{longexpression}), the last part having $A_{ia}^{\mu_{n+2}}$ vanishes thanks to Eq.~(\ref{AA}).
The remaining term gives Eq.~(\ref{2sspre}), once $\partial B$ is rewritten into $\cal B$ using Eq.~(\ref{calAdef}).

\section{Absence of invertible field transformation with derivatives for one field case}
\label{App:one-field}

In this appendix, we show that there is no invertible field transformation of one field involving up to first derivative by use of the conditions obtained for two field case. We consider the following transformation, 
\begin{align}
\phi_1 &= \bar \phi_1(\psi_1, \partial_\alpha \psi_1, x^\mu), \\
\phi_2 &= \psi_2,
\label{eq:one_field}
\end{align}
which essentially represents one-field transformation with derivatives. Then, the first degeneracy condition~(\ref{aVU}) is satisfied as follows,
\begin{equation}
A^\mu_{ia}
= \frac{\partial \phi_i}{\partial\left(\partial_\mu \psi_a\right)}
= \left( \begin{array}{cc}
\frac{\partial \phi_1}{\partial\left(\partial_\mu \psi_1\right)}  & 0 \\
0 & 0 \\
\end{array} \right)
= \frac{\partial \phi_1}{\partial\left(\partial_\mu \psi_1\right)} U_a V_i~,
\label{eq:one_first_cond}
\end{equation}
where $\frac{\partial \phi_1}{\partial\left(\partial_\mu \psi_1\right)}$ is assumed to be non-zero and
\begin{equation}
U_a = (1,0) = V_i,
\quad
m_a = (0,-1) = n_i.
\end{equation}
One the other hand, the second degeneracy condition~(\ref{nBm}) is never satisfied as follows,
\begin{gather}
B_{ia}
= \frac{\partial \phi_i}{\partial \psi_a}
= \left( \begin{array}{cc}
\frac{\partial \phi_1}{\partial \psi_1} & 0 \\
0 & 1 \\
\end{array} \right), \\
n_i B_{ia} m_a = 1 \ne 0.
\end{gather}
Note that one can also check easily that third conditions (\ref{nBU}) are also violated because $n_i B_{ia}U_a = 0$ and $\left( V_j B_{jb} -  a^\beta \partial_\beta U_b \right) m_b = 0$ in this case. Thus, there is no invertible field transformation of one field involving first derivatives.
A similar argument applies also to a case with more derivatives, $\phi_1 = \bar \phi_1(\psi_1, \partial_\alpha \psi_1, \partial_\alpha\partial_\beta \psi_1, \ldots ; x^\mu)$, and we can show that it can never be invertible.

\section{Inverse function theorem applied directly to the functional space, and implicit function theorem to finite-dimensional subspaces
}
\label{App:implicit-function}

In our proof in sections~\ref{sec:necessary} and \ref{Sec:Suf}, we do not apply the inverse function theorem to the mapping from $\psi(x)$ to $\phi(x)$, 
but instead use the implicit function theorem to the mapping from $(\psi, \p \psi, \ldots)$ to $(\phi, \p\phi, \ldots)$. 
This is because the former indeed does not work in the cases with derivatives. 
In this appendix, the reason why it is not applicable is explained in section~\ref{inappIFT} and 
show the idea of our proof in section~\ref{idea}.

\subsection{Inapplicability of inverse function theorem to the functional space}\label{inappIFT}

The inverse function theorem can be applied to the mapping from a Banach space $X$ to another Banach space $Y$. 
Functional spaces that are discussed in physics are mainly Hilbert spaces, and thus 
one might naively think that the inverse function theorem is directly applicable to a field redefinition with derivatives from $\psi(x)$ to $\phi(x)$ by analyzing the whole spacetime points simultaneously. 
However, if derivatives of fields are involved in the field redefinition, the inverse function theorem can not be applied to the functional space. 
This means that the linearized analysis in the functional space does not give correct statements. 
Let us see that the inverse function theorem in functional spaces does not work in the cases with derivatives.

We have the condition for the degeneracy of the coefficients of the highest order derivative, 
\begin{equation}
\det (A^\mu_{ia} \xi_\mu) = 0. 
\label{Amu}
\end{equation}
In the (wrong) linear analysis, this is obtained as an equation which has to be satisfied for a fixed value of $\psi_a$. 
However, we know from the analysis of the necessary condition 
discussed in section~\ref{sec:necessary}
that 
Eq.~\eq{Amu} should be interpreted as an identity, {\it i.e.} the equality has to be satisfied for any $\psi_a$. 
Since the sufficient condition, which can be obtained from the inverse function theorem, should be stronger than the necessary condition,
Eq.~\eq{Amu} should be obtained as an identity (or replaced by a stronger condition).

We shall see in detail why the inverse function theorem is not applicable directly to the field redefinition with derivatives. 
For this purpose, let us carefully inspect the statement of the inverse function theorem for a mapping between Banach spaces \cite{Luenberger}:

\vspace{2mm}

\noindent
{\it 
Let $X$ and $Y$ be Banach spaces.
Let $U$ be an open neighborhood of a point $x_0\in X$
and $F$ be a continuously differentiable mapping from $U$ to $Y$, $F:U\to Y$.
Suppose that there exists the Fr\'echet derivative $dF_0$ at $x_0 \in U$ which gives a bounded linear isomorphism that maps $U$
onto an open neighborhood of $F(x_0) \in Y$.
Then, there exists an open neighborhood $U'$ of $x_0\in X$ and $V$ of $F(x_0) \in Y$ and 
a continuously differentiable map $G$ from $V$ to $U'$, $G: V \to U'$ satisfying $F(G(y))=y$.
}

\vspace{2mm}

The important point is that, in the application of the theorem,
a mapping is required to be continuously differentiable.
To see the argument clearly, we consider an example given by
\begin{equation}
\phi_1 = \psi_1 \dot \psi_2 + \psi_1 ~,
\qquad \phi_2 = \dot \psi_1 + \psi_2~.
\label{RD}
\end{equation}
We denote this mapping as $\Phi: \psi_a(x) \to \phi_i(x)$.
Let us apply the linear analysis to $\Phi$ (although it gives an incorrect result).
We linearize Eq.~\eqref{RD} as
\begin{equation}
\delta\phi_1 = \delta\psi_1 \dot \psi_2 + \psi_1 \dot{ \delta \psi_2} + \delta \psi_1~, 
\qquad \delta\phi_2 = \dot {\delta\psi_1} + \delta\psi_2~.
\end{equation}
If we consider $\psi_1(x)=\psi_2(x)=0$ for any $x$, a point in the phase space, the above equations become
\begin{equation}
\delta\phi_1 =  \delta \psi_1~, 
\qquad \delta\phi_2 = \dot {\delta\psi_1} + \delta\psi_2~,
\end{equation}
and they can be solved uniquely for $\delta \psi_1, \delta\psi_2$ as
\begin{equation}
\delta \psi_1= \delta\phi_1~, 
\qquad \delta\psi_2= \delta\phi_2 - \dot {\delta\phi_1} ~ .
\end{equation}
Since the linearized equation is uniquely solved, one might think the inverse function theorem is applicable, 
but it is not true. 
This is because, if the inverse function theorem is applicable,
the mapping $\Phi$ must be invertible also for $\psi_1(x)=\epsilon(\neq 0)$ and $\psi_2(x)=0$, which is in the neighborhood of $\psi_1(x)=\psi_2(x)=0$, for sufficiently small $\epsilon$.
However, it is not the case because the linearized equations for $\psi_1(x)=\epsilon$, $\psi_2(x)=0$ become
\begin{equation}
\delta\phi_1 = \epsilon \dot{ \delta \psi_2} + \delta \psi_1 ~ , 
\qquad \delta\phi_2 = \dot {\delta\psi_1} + \delta\psi_2 ~ .
\end{equation}
These equations cannot be solved uniquely for $\delta \psi_1, \delta\psi_2$, hence the transformation $\Phi$ is not invertible.

What was wrong in this example? 
Actually, derivative of the mapping $\Phi\bigl(\psi_a(x)\bigr)$ (that is, the field redefinition \eq{RD}) can not be obtained in the linearized analysis. 
The Fr\'{e}chet derivative of $\Phi$ is defined as follows. 

\vspace{2mm}

\noindent
{\it
If there exists a bounded linear operator $A$ satisfying
\begin{equation}
\lim_{|| \delta \psi_a(x)  ||_M \to 0}
 \frac{||  \Phi\bigl(\psi_a(x) + \delta \psi_a(x) \bigr)- \Phi\bigl(\psi_a(x) \bigr) -A \, \delta \psi_a(x) ||_M  }
 { || \delta \psi_a(x)  ||_M }=0,
\end{equation}
the Fr\'{e}chet derivative is defined as 
\begin{equation}
D \Phi \bigl(\psi_a(x) \bigr) = A. 
\end{equation}
Otherwise, the Fr\'{e}chet derivative does not exist. 
Here, $M$ is an open region in spacetime that we consider and $||\cdot||_M$ is a norm for the phase space 
covered by $\phi_i(x)$ or $\psi_a(x)$.
}

\vspace{2mm}

For the existence of the Fr\'{e}chet derivative that is necessary to establish the invertibility, the linear part of
\begin{equation}
  \Phi\bigl(\psi_a(x) + \delta \psi_a(x) \bigr)- \Phi\bigl(\psi_a(x) \bigr) 
= \left(\delta\psi_1 \dot \psi_2 + \psi_1 \dot{ \delta \psi_2} + \delta \psi_1 \dot{ \delta \psi_2} + \delta \psi_1, 
\dot {\delta\psi_1} + \delta\psi_2
\right)
\label{Frechet}
\end{equation}
must be close to
$A\delta \psi_a(x)$, where $A$ is a {\it bounded} linear operator. 
One may expect that the linear part of \eq{Frechet} can be approximated by $A\delta \psi_a(x)$. 
However, Eq.~\eqref{Frechet} includes derivatives of $\delta\psi$, which may not be bounded. 
Hence, the inverse function theorem is not applicable.

The inverse functional theorem roughly means that, if it is applicable, the linear analysis works well.
However, in the application to an infinite dimensional space, such as a functional space, 
the existence of the Fr\'{e}chet derivative is required. 
Without confirmation of it, the result of the invertibility by the linear analysis is untrustable and, 
in the case of the field redefinition with derivative, it generically does not work.

\subsection{Our idea: Application of the implicit function theorem to the finite-dimensional subspaces}
\label{idea}

To avoid complication due to the derivatives involved in the mapping, 
we take a different approach
by working in the function space spanned by  $\phi_i(x_0)$, $\partial_\mu \phi_i(x_0)$, $\partial_\mu  \partial_\nu \phi_i(x_0)$, $\ldots$ as follows.
The field redefinition (\ref{redef}) from $\psi_a$ to $\phi_i$ is obviously unique; 
Eq.~(\ref{redef}) shows that $\psi_a$ is uniquely obtained from a fixed $\phi_i$.
Hence, if also the mapping from $\phi_i$ to $\psi_a$ is unique, the field redefinition becomes invertible. 

Let us give $\phi_i$ in an open region of spacetime and  then its derivatives $\partial_\mu \phi_i$, $\partial_\mu  \partial_\nu \phi_i$, $\ldots$ are uniquely obtained. 
Thus, 
if we can show that $\psi_a$ is uniquely fixed for given $\phi_i$ and its derivatives $\partial_\mu \phi_i$, $\partial_\mu  \partial_\nu \phi_i$, $\ldots$~, we can say the inverse mapping is unique.
To show it, we use implicit function theorem.

The problem appearing in the direct application of the inverse function theorem in section~\ref{inappIFT} stems from that 
the dimension of the space is infinite. 
Even if we use the implicit function theorem, we have a similar problem.
However, if we use equations shown in section~\ref{Sec:Suf}, 
we can consider the implicit function theorem at each point of spacetime separately, 
{\it i.e.} for $\psi_a(x_0)$ with fixed $\phi_i(x_0)$ and its derivatives $(\partial_\mu \phi_i(x_0),\partial_\mu  \partial_\nu \phi_i(x_0),\ldots)$ at each spacetime point $x_0$.
Note that now we (are trying to) apply the implicit function theorem to  $(\phi_i(x_0);\phi_i(x_0) \partial_\mu \phi_i(x_0),\partial_\mu  \partial_\nu \phi_i(x_0),\ldots)$ with fixed $x_0$, and thus the dimension of space is finite. 
(Here, we consider the case where the number of derivative of $\phi$ is finite.)

Since the dimension of space is finite, we do not bother the infiniteness of the dimension and
the implicit function theorem  can be easily applied. 
Note that the neighborhood in the implicit function theorem here means that for the values of $\phi_i(x_0)$, $\partial_\mu \phi_i(x_0)$, $\partial_\mu  \partial_\nu \phi_i(x_0)$, $\ldots$, not spacetime point. 
Hence, the neighborhood or ``locality'' in our argument means  a local region of functional space, 
where the deviations of $\phi_i(x)$, $\partial_\mu \phi_i(x)$, $\partial_\mu  \partial_\nu \phi_i(x)$, $\ldots$ 
from a reference value should be small. 
It would be acceptable in physics, which usually has the UV cutoff scale.

Note that ``neighborhood'' in the implicit (or inverse) function theorem is not necessarily small. 
Let us take a simple example $f(x)=x^2$.
The inverse function theorem is applicable except $x=0$. 
At $x=3$, one might take an open neighborhood $2<x<4$.
Then, we take the inverse function theorem again near $x=2$ and the ``neighborhood'' can be extended finally to $0<x$. 
This extension of ``neighborhood'' is general and it is done just before the condition is violated. 

\section{Setting $U_a=(0,1)$ in $\phi_i = \phi_i\left(\psi_a, U_a(\psi_a)\partial_\mu \psi_a\right)$}
\label{App:Setting-U}

In this appendix, we consider a transformation in which derivative of $\psi_a$ appears only in a combination $U_a(\psi_a)\partial_\mu \psi_a$, that is,
\begin{equation}
\phi_i = \phi_i\left(\psi_a, U_a(\psi_a)\partial_\mu \psi_a\right)~,
\label{trsfUdpsi}
\end{equation}
and show that $U_a(\psi_a)$ can be set to a constant vector $U_a = (0,1)$ without loss of generality by a field transformation $\psi_a = \psi_a (\tilde \Psi_a)$.
This technique is used in section~\ref{sec:examples} to simplify examples of invertible transformations.

\subsection{Field transformation to set $U_a=(0,1)$}
\label{App:Setting-U_sub}

As a first step to set $U_a(\psi_a)$ to a constant vector, 
we rewrite this transformation as
\begin{equation}
\phi_i = \phi_i\bigl(\psi_a, \tilde U_a(\psi_a)\partial_\mu \psi_a\bigr)~,
\label{tildetrsf}
\end{equation}
where
\begin{equation}
\tilde U_a\equiv c(\psi_a)U_a(\psi_a)
\label{rescaling}
\end{equation}
is a local rescaling of $U_a$ such that $\tilde U_a$ is an irrotational vector in the $\psi_a$ space, that is,
\begin{equation}
 \frac{\partial \tilde U_1}{\partial \psi_2} - \frac{\partial \tilde U_2}{\partial \psi_1} = 0~.
\label{irrotational}
\end{equation}
This $\tilde U_a$ can be obtained by choosing the rescaling factor $c(\psi_a)$ appropriately.
We show the construction method of $c(\psi_a)$ and $\tilde U_a(\psi_a)$ in the next section.

When (\ref{irrotational}) is satisfied, 
due to the Poincar\'e lemma
there exists a scalar function $\Psi_1(\psi_k)$ satisfying
\begin{equation}
 \tilde U_a = \frac{\partial \Psi_2}{\partial \psi_a}~.
\end{equation}
Let us also introduce another scalar function $\Psi_1(\psi_a)$ that is functionally independent from $\Psi_1(\psi_a)$ ({\it i.e.} $\det(\partial\Psi_a/ \partial\psi_b)\neq 0$).
Then there exists are one-to-one mapping between $\psi_a$ and $\Psi_a$, and
the transformation (\ref{tildetrsf}) may be expressed in terms of $\Psi_a$ as 
\begin{equation}
\phi_i = \phi_i\bigl(\Psi_a, \partial_\mu \Psi_2\bigr)~.
\label{tildetrsf2}
\end{equation}
This is equivalent to setting $U_a = (0,1)$ in the transformation (\ref{trsfUdpsi}).
Hence, we may impose $U_i=(0,1)$ without loss of generality in the transformation~(\ref{trsfUdpsi}), as long as there exists a rescaling $c(\psi_i)$ satisfying the condition~(\ref{irrotational}).

\subsection{Finding irrotational $\tilde U_i$}
\label{app:irrotationalU}

In the argument above, it is crucial that 
there exists a rescaling (\ref{rescaling}) that makes $\tilde U_a$ an irrotational vector satisfying (\ref{irrotational}).
We show how to find such a rescaling in this section.

Using Eq.~(\ref{rescaling}), the condition~(\ref{irrotational}) can be rewritten as
\begin{equation}
  U_1 \frac{\partial  c}{\partial \psi_2}
- U_2 \frac{\partial  c}{\partial \psi_1}
+ c\left(
\frac{\partial U_1}{\partial \psi_2}
- \frac{\partial U_2}{\partial \psi_1}
\right)
= 0~.
\label{ceq}
\end{equation}
This equation is a first-order partial differential equation for $\log c(\psi_1, \psi_2)$, and it can be solved a least locally once an appropriate boundary condition for $c$ is given.
For example, when $U_1\neq 0$ we may solve Eq.~(\ref{ceq}) as an evolution equation in the ``time'' direction $\psi_2$ 
for an initial condition given by $c=1$ on a $\psi_2=\text{constant}$ line on the $\psi_a$ space,
regarding $\psi_1$ as the ``spatial'' coordinate.

\section{Disformal transformation with higher derivatives for the FRW ansatz}
\label{app:gendisformal}

In section~\ref{sec:no-go} we examined a generalization of the disformal transformation to introduce the second derivative of the scalar field $\nabla\nabla\chi$, and found it non-invertible unless such a second-derivative term is absent. In this appendix we consider a further generalization of this transformation, and find that there may be an invertible disformal transformation with higher derivatives if the metric is limited to the FRW type.
Note that the investigation of the FRW type gives a necessary condition. 
While the violation of invertibility in the FRW subspace shows that of full space of metric, 
the establish in subspace does not result in that in full space.

\subsection{Generalized disformal transformation with second derivatives}

Let us consider the following generalization of the disformal transformation~\cite{Minamitsuji:2021dkf,Alinea:2020laa}
\begin{equation}
	\tilde{g}_{\mu\nu} = \mathcal{F}_0 g_{\mu\nu}+ \mathcal{F}_1 \chi_\mu\chi_\nu+ \mathcal{F}_2 \chi_{\mu\nu} +  \mathcal{F}_3 \chi_{\left(\mu\right.}X_{\left.\nu\right)} 
	+ \mathcal{F}_4 X_\mu X_\nu + \mathcal{F}_5 \chi_\mu^\alpha\chi_{\nu \alpha},
	\label{CT}
\end{equation}
where $\mathcal{F}_i$ depends on $X$, $\mathcal{B}$, $\mathcal{Y}$, $\mathcal{Z}$ and $\mathcal{W}$, which are scalar quantities involving up-to square of the second derivative of the scalar field $\nabla\nabla\chi$, that is, 
\begin{equation}
X= \chi^\mu\chi_\mu,
\quad
\mathcal{B} = \Box\chi,
\quad
\mathcal{Y} = \chi^\mu X_\mu,
\quad
\mathcal{Z} = X_\mu X^\mu,
\quad
\mathcal{W}= \chi^{\mu\nu}\chi_{\mu\nu}.
\end{equation}
The subscripts denote covariant derivatives (e.g., $\chi_\mu = \nabla_\mu \chi,\; \chi_{\mu\nu}=\nabla_\mu \nabla_\nu \chi$).

Let us take the FRW  homogeneous ansatz for the metric $g_{\mu\nu}$,
\begin{equation}
g_{\mu\nu}dx^\mu dx^\nu = -\mn(t) dt^2 + \ma(t)d{\bf x}^2, \quad \chi=\chi(t).
\label{FRW}
\end{equation}
For this ansatz we have,
\begin{equation}
\begin{aligned}
X & = -\frac{\dot\chi^2}{\mn},\\
\mathcal{B} & =\frac{1}{2\mn}\left( \dot{\chi}\frac{\dot{\mn}}{\mn} -3\dot{\chi}\frac{\dot{\ma}}{\ma} - 2\ddot{\chi} \right) = \frac{\mathcal{Y}}{2X}-\frac{3\dot\chi}{2\mn}\frac{\dot\ma}\ma,\\
\mathcal{Y} & =-\frac{\dot\chi^2}{\mn^2}\left( \dot{\chi}\frac{\dot\mn}{\mn} - 2\ddot{\chi} \right),\\
\mathcal{Z} & =\frac{\dot\chi^2}{\mn^3}\left( -\dot{\chi}^2\frac{\dot{\mn}^2}{\mn^2} + 4\dot{\chi}\ddot{\chi}\frac{\dot{\mn}}{\mn} - 4 \ddot{\chi}^2 \right)
	=-\frac{\dot\chi^2}{\mn^3}\left( \dot\chi\frac{\dot\mn}\mn-2\ddot\chi \right)^2 = \frac{\mathcal{Y}^2}X,\\
\mathcal{W} & = \frac{1}{\mn^2}\left( \frac{\dot{\chi}^2}4 \frac{\dot{\mn}^2}{\mn^2} + \frac{3\dot{\chi}^2}{4}\frac{\dot{\ma}^2}{\ma^2} - \dot{\chi}\ddot{\chi}\frac{\dot{\mn}}{\mn} +\ddot{\chi}^2 \right)
= \left(\frac{\mathcal{Y}}{2X}\right)^2 +\frac13\left( \frac{\mathcal{Y}}{2X} -\mathcal{B}\right)^2.
\end{aligned}
\end{equation}
Hence, for the FRW ansatz, $\mathcal{Z}$ and $\mathcal{W}$ are expressed by $X,\; \mathcal{B}$ and $\mathcal{Y}$ and not independent.
Evaluating also the terms appearing in Eq.~(\ref{CT}), we find that the metric after the transformation $\tilde g$ is expressed by the FRW ansatz
\begin{equation}
\tilde g_{\mu\nu} = \tilde \mn(t) dt^2 + \tilde \ma(t) dx^2 ,
\end{equation}
and $\tilde \mn, \tilde \ma$ are given by
\begin{equation}
\begin{aligned}
\tilde\mn &=  \mn \left(\mathcal{F}_0 + \mathcal{F}_1X +\mathcal{F}_2 \frac{\mathcal{Y}}{2X} +\mathcal{F}_3  {\mathcal{Y}}+\mathcal{F}_4 \frac{\mathcal{Y}^2}X + \mathcal{F}_5\left(\frac{\mathcal{Y}}{2X} \right)^2 \right)
:=\mn F_\mn(X,\mathcal{B},\mathcal{Y})
\\
\tilde \ma &=  \ma \left(\mathcal{F}_0 -\mathcal{F}_2 \frac13 \left( \frac{\mathcal{Y}}{2X} -\mathcal{B}\right)  + \mathcal{F}_5\frac19 \left( \frac{\mathcal{Y}}{2X} -\mathcal{B}\right)^2 \right)
:=\ma F_\ma(X,\mathcal{B},\mathcal{Y}).
\end{aligned}
\label{CT2}
\end{equation}
This result implies that the generalized disformal transformation (\ref{CT}) is completely governed by the functions $F_\mn(X,\mathcal{B},\mathcal{Y})$ and $F_\ma(X,\mathcal{B},\mathcal{Y})$ for the FRW ansatz.

\subsection{The general disformal transformation reduced on the FRW spacetime}

The transformation (\ref{CT2}) involves only $\mn, \ma$ and their first derivatives, hence its invertibility can be analyzed within the framework explained in section~\ref{sec:necessary}.
Based on it, let us derive the invertibility conditions for the transformation (\ref{CT2}) below.

Since the transformation (\ref{CT2}) involves $\dot \mn, \dot \ma$, $A_{ia}$ must be degenerate for the invertibility:
\begin{equation}
A_{ia}
=
\begin{pmatrix}
\partial \tilde\mn / \partial {\dot\mn}
&
\partial \tilde\mn / \partial {\dot\ma}
\\
\partial \tilde\ma / \partial {\dot\mn}
&
\partial \tilde\ma / \partial {\dot\ma}
\end{pmatrix}
=
\frac12
\sqrt{\frac{-X}{\mn}}
\begin{pmatrix}
F_{\mn,\mathcal{B}}
+
2X F_{\mn,\mathcal{Y}}
&
-\frac{3\mn}{\ma}F_{\mn,\mathcal{B}}
\\
\frac{\ma}{\mn}
\left(
F_{\ma,\mathcal{B}}
+
2X F_{\ma,\mathcal{Y}}
\right)
&
-3F_{\ma,\mathcal{B}}
\end{pmatrix}
\end{equation}
\begin{equation}
\det A = \frac{3X^2}{2\mn} \left(
F_{\ma,\mathcal{B}} F_{\mn,\mathcal{Y}}
- F_{\mn,\mathcal{B}} F_{\ma,\mathcal{Y}}
\right) = 0.
\label{detA=0_gen}
\end{equation}
Below we assume that the rank of the matrix $A_{ia}$ is 1. When the rank of $A_{ia}$ is zero, all the components of $A_{ia}$ vanishes and then the transformation (\ref{CT2}) does not involve $\dot \mn$ and $\dot \ma$. In this case, the invertibility condition is simply given by $\det B_{ia} \neq 0$.

When the degeneracy condition~(\ref{detA=0_gen}) is satisfied, there exists zero eigenvectors $n_i, m_a$ and their dual vectors $V_i, U_a$ satisfying
\begin{equation}
n_i A_{ia} = 0 = A_{ia} m_a,
\quad
n_i n_i = 1 = m_a m_a,
\quad
n_i = \epsilon_{ij}V_j,
\quad
m_a = \epsilon_{ab}U_b.
\end{equation}
They are explicitly given as
\begin{equation}
\begin{gathered}
n_i = N^{-1} \left(
F_{\ma,\mathcal{B}}, -\frac{\mn}{\ma} F_{\mn,\mathcal{B}}
\right),
\quad
m_a = M^{-1} \left(
\frac{3\mn}{\ma} F_{\mn,\mathcal{B}},
2X F_{\mn,\mathcal{Y}} + F_{\mn,\mathcal{B}}
\right),
\\
V_i = N^{-1} \left(
-\frac{\mn}{\ma} F_{\mn,\mathcal{B}},
-F_{\ma,\mathcal{B}}
\right),
\quad
U_a = M^{-1} \left(
2X F_{\mn,\mathcal{Y}} + F_{\mn,\mathcal{B}},
-\frac{3\mn}{\ma} F_{\mn,\mathcal{B}}
\right),
\end{gathered}
\end{equation}
where $M$ and $N$ are normalization coefficients given by
\begin{equation}
M = 
\sqrt{
\left(
F_{\mn,\mathcal{B}} + 2 X F_{\mn,\mathcal{Y}}
\right)^2
+\left(
3\mn F_{\mn,\mathcal{B}} / \ma
\right)^2
},
\quad
N = 
\sqrt{
F_{\ma,\mathcal{B}}^2
+\left(
\mn F_{\mn,\mathcal{B}} / \ma
\right)^2
}.
\label{MNdef}
\end{equation}
We assume $M, N\neq 0$ below, and discuss the case where either $M$ or $N$ vanishes separately.

Using these expressions, the second degeneracy condition (\ref{nBm}) in the invertibility conditions is evaluated as
\begin{equation}
0=n_iB_{ia}m_a
=
\frac{1}{MN}
\frac{\mn F_{\mn,\mathcal{B}}}{ \ma }
\left[
3 F_{\mn} F_{\ma,\mathcal{B}}
- F_{\ma} 
\left(
F_{\mn,\mathcal{B}}
+ 2 X  F_{\mn,\mathcal{Y}}
\right)
+ 3 X 
\left(
F_{\ma,X} F_{\mn,\mathcal{B}}
- F_{\ma,\mathcal{B}} F_{\mn,X}
\right)
\right]~,
\label{nBm=0_gen}
\end{equation}
where we eliminated $F_{\ma,\mathcal{Y}}$ using the first degeneracy condition (\ref{detA=0_gen}).

The non-degeneracy conditions~(\ref{nBU}) in the invertibility conditions are given as follows.
\begin{align}
0\neq n_i B_{ia} U_a
&=
\frac{MF_\ma}{3N}
~,
\label{firstnondeg}
\end{align}
where we used the first and second degeneracy conditions (\ref{detA=0_gen}), (\ref{nBm=0_gen}) to simplify the expression.
The other non-degeneracy condition can be derived after some calculations as
\begin{equation}
0\neq \left(V_i B_{ia} - a \dot U_a\right)m_a
=
\frac{3 N}{M}
\biggl[
- \left(-X\right)^{3/2}
F_{\mn,\mathcal{B}}
\left(\frac{F_{\mn,\mathcal{Y}}}{F_{\mn,\mathcal{B}}}\right)'
- F_{\mn}
+ X F_{\mn,X}
+ \mathcal{B} F_{\mn,\mathcal{B}}
+ \mathcal{Y} F_{\mn,\mathcal{Y}}
\biggr],
\label{secondnondeg}
\end{equation}
where a prime ($'$) denotes a derivative with respect to the proper time, that is, $f'\equiv \mn^{-1/2} \dot f$.
To simplify this equation, we used the degeneracy conditions (\ref{detA=0_gen}), (\ref{nBm=0_gen}) and their $t$ derivatives, that is, we assumed the degeneracy conditions (\ref{detA=0_gen}), (\ref{nBm=0_gen}) are satisfied identically at any $t$.

To summarize, the transformation (\ref{CT2}) becomes invertible when the conditions (\ref{detA=0_gen}), (\ref{nBm=0_gen}), (\ref{firstnondeg}) and (\ref{secondnondeg}) are satisfied.
These conditions should be regarded as necessary conditions for the invertibility of the generalized disformal transformation (\ref{CT}) for a general metric, because the above results are derived only for the FRW ansatz (\ref{FRW}), in which the degrees of freedom of the metric is reduced to two functions  $\mn(t), \ma(t)$ depending only on $t$. 
The invertibility conditions for a general metric should encompass the conditions (\ref{detA=0_gen}), (\ref{nBm=0_gen}), (\ref{firstnondeg}) and (\ref{secondnondeg}), while they may contain more stringent conditions in general. 
An obvious next step is to derive the invertibility conditions for a more general metric, and also it would be interesting to construct examples of invertible transformation based on the invertibility conditions obtained above.
We reserve those issues for future works. 

Let us briefly mention the case that either $M$ or $N$ given by Eq.~(\ref{MNdef}) vanishes.
In this case, it turns out that the degeneracy conditions imply that $F_{\mn,\mathcal{B}}=F_{\mn,\mathcal{Y}}=F_{\ma,\mathcal{B}}=0$, that is,
$F_\mn = F_\mn(X)$ and $F_\ma = F_\ma(X, \mathcal{Y})$.
For these functions it follows that
\begin{equation}
A_{ia}
=
-\left(\frac{-X}{\mn}\right)^{3/2}
\begin{pmatrix}
0
&
0
\\
\ma F_{\ma,\mathcal{Y}}
&
0
\end{pmatrix}
\end{equation}
and $V_i=(0,1)$, $U_a=(1,0)$. 
It can be shown that
the first and second degeneracy conditions are automatically satisfied, and the non-degeneracy conditions are given by
\begin{equation}
F_\mn - X F_{\mn,X}\neq 0,
\qquad
F_\ma \neq 0.
\end{equation}

\end{document}